\documentclass[twocolumn,3p,times]{elsarticle}

\usepackage{graphicx}
\usepackage{geometry}
\usepackage{booktabs}
\usepackage{array}
\usepackage{epic,eepic,amsmath}
\usepackage{graphics}
\usepackage{epsfig}
\usepackage{amsmath}
\usepackage{bm}
\usepackage{amssymb}
\usepackage{natbib}

\usepackage{color}
\usepackage{amssymb}
\usepackage{amsmath}
\usepackage{graphicx}
\usepackage{colortbl}

\usepackage{amssymb}
\usepackage{enumerate}
\usepackage{amsthm,amsfonts,amsmath}
\usepackage{bm}
\usepackage{graphicx}
\usepackage{graphics,epsf,psfrag,pifont,dsfont,bbold}
\usepackage{siunitx}
\usepackage{color}
\usepackage{xspace}
\usepackage{accents}

\definecolor{light-gray}{gray}{0.6}

\renewcommand{\vec}[1]{\ensuremath{\mathbf{#1}}}

\newcommand{\pdev}[1]{\ensuremath{\partial_{#1}}}

\newcommand*{\vcenteredhbox}[1]{\begingroup       
\setbox0=\hbox{#1}\parbox{\wd0}{\box0}\endgroup}   

\newcommand*{\hcenteredhbox}[1]{\begingroup       
\setbox0=\vbox{#1}\parbox{\wd0}{\box0}\endgroup}   

\begin{document}

\begin{frontmatter}

\title{Lattice Boltzmann simulations of droplet dynamics in time-dependent flows \tnoteref{postprint}}
\author[address1]{F. Milan \corref{email}} 
\author[address2]{M. Sbragaglia}
\author[address2]{L. Biferale}
\author [address3]{F. Toschi}

\tnotetext[postprint]{Postprint version of the article published in Eur. Phys. J. E (2018) \textbf{41}: 6 doi: 10.1140/epje/i2018-11613-0}
\cortext[email]{\texttt{felix.milan@roma2.infn.it}}           

\address[address1]{Department of Physics and INFN, University of ``Tor Vergata'', Via della Ricerca Scientifica 1, 00133 Rome, Italy and Department of Applied Physics, Eindhoven University of Technology, Eindhoven 5600 MB, The Netherlands}
\address[address2]{Department of Physics and INFN, University of ``Tor Vergata'', Via della Ricerca Scientifica 1, 00133 Rome, Italy}
\address[address3]{Department of Applied Physics and Department of Mathematics and Computer Science, Eindhoven University of Technology, Eindhoven 5600 MB, The Netherlands and CNR-IAC, Rome I-00185, Italy}

\begin{abstract}
We study the deformation and dynamics of droplets in time-dependent flows using 3D numerical simulations of two immiscible fluids based on the lattice Boltzmann model (LBM).  Analytical models are available in the literature, which assume the droplet shape to be an ellipsoid at all times (P.L. Maffettone, M. Minale, J. Non-Newton. Fluid Mech \textbf{78}, 227 (1998); M. Minale, Rheol. Acta \textbf{47}, 667 (2008)). Beyond the practical importance of using a mesoscale simulation to assess ``ab-initio'' the robustness and limitations of such theoretical models, our simulations are also key to discuss - in controlled situations - some relevant phenomenology related to the interplay between the flow time scales and the droplet time scales regarding the ``transparency" transition for high enough shear frequencies for an external oscillating flow. This work may be regarded as a step forward to discuss extensions towards a novel DNS approach, describing the mesoscale physics of small droplets subjected to a generic hydrodynamical strain field, possibly mimicking the effect of a realistic turbulent flow on dilute droplet suspensions.
\end{abstract}

\end{frontmatter}

\thispagestyle{empty}

\section{Introduction}
\label{sec:intro}

Mesoscale numerical simulations, and in particular the Lattice Boltzmann methods (LBM), are useful computational tools for the description of a wide range of multiscale problems, distinctly characterized by a coupling between the physics at microscales (e.g. interface dynamics, polymers, thermal fluctuations) and the resulting manifestations on the hydrodynamical flow at large scales (e.g. flow-structure coupling, capillary fluctuations etc). Due to their built-in properties, LBM live at mesoscales and can approach the hydrodynamical description via a coarse grained procedure starting from probability distribution functions at the kinetic level \cite{Benzi92,Succi01}. The latter, in turn, can be efficiently enriched with various microscopic ingredients, such as non-ideal effects \cite{Sbragaglia12}, coupling with polymer micro-mechanics \cite{Onishi1}, thermal fluctuations \cite{Varnik11}. This makes LBM very successful in simulating the physics of fluids over a very broad range of scales. The multiscale problem of interest in this paper is the fluid dynamics of an emulsion, i.e. a collection of small deformable droplets dispersed in a solvent immiscibile fluid: droplets can deform under the action of an imposed flow and can interact with neighboring droplets, they provide a back-reaction on the solvent component and ultimately determine the complex flowing properties of the emulsion at large scales. Far from being only an interesting multiscale physical problem, it also finds a variety of applications in industrial and engineering processes \cite{Flumerfelt72}. In an attempt to disentangle key physical ingredients out of this complex scenario, the simplest problem to analyze is the dynamics and deformation of a single droplet under the influence of an externally imposed flow. The literature on single droplet deformation is vast, especially when dealing with laminar flows: following the pioneering work by Taylor \cite{Taylor32}, the deformation properties of a droplet have been extensively studied and reviewed \cite{Greco02,GuidoRev}. Existing studies address the effects introduced by the nature of the flow \cite{GuidoSimeoneGreco,Sibilloetal06,Premnath,Tome}, the effects of confinement \cite{Minale08,ShapiraHaber90}, as well as the effects introduced by the complex non-Newtonian nature of the bulk fluids \cite{MaffettoneGreco04,MaffettoneGrecoSimeoneGuido,GuptaSbragaglia2014}. Exact analytical approaches are typically limited to ``small'' deformation assumptions, i.e. perturbative results. Extensions to time-dependent laminar flows have also been carried out \cite{Greco02b}. From the theoretical side, a popular model has been developed by Minale \& Maffettone \cite{MaffettoneMinale98} (hereafter MM). The MM model is characterized by three key ingredients: firstly, the droplet deformation, which is parametrized by the Capillary number $\mbox{Ca}$, secondly, the viscous ratio $\chi=\eta_{\tiny \mbox{d}}/\eta_{\tiny \mbox{s}}$, where $\eta_{{\tiny \mbox{d}},{\tiny \mbox{s}}}$ is the dynamic viscosity of the droplet (d) or solvent (s) phase, and thirdly, the (imposed) time-dependent strain matrix which is the input for the model and changes with the flow topology. The model assumes that the droplet is an ellipsoid at all times \cite{Guido98,Stone}, and is constructed to recover the perturbative results on droplet deformation at small $\mbox{Ca}$ (e.g. Taylor's result \cite{Taylor32}) in the presence of a steady flow. The MM model has the key advantages to allow for time dynamics and also to extend the description of droplet deformation beyond the limits of applicability of perturbation theories \cite{Taylor32}, hence it has also been used to characterize the critical  $\mbox{Ca}$ for which droplet break-up occurs \cite{MaffettoneMinale98}. Following the MM model, a whole class of ``ellipsoidal'' models have been introduced with further enrichments to account for a variety of other effects, including viscoselasticity \cite{Verhulst09a,Verhulst09b,AggarwalSarkar08,Mukherjee10}, confinement\cite{GuptaSbragaglia2014,Cardinaels11,Guido11,Cardinaels09,Cardinaels10,Cardinaelsetal11b,GuptaSbragagliaScagliarini2015,CardinaelsMoldenaers10} and matching with more
 refined perturbative results at small $\mbox{Ca}$ \cite{Sibilloetal06,MaffettoneGreco04,MaffettoneGrecoSimeoneGuido,Yue04}. A detailed review on the topic can be found in \cite{Minale10b}. These ellipsoidal models become particularly useful when studying the properties of a single droplet under the influence of turbulent fluctuations \cite{Njobuenwu2015,Biferale2014,Spandan2016}. Depending on the characteristic size of the droplet, turbulent fluctuations can provide either inertial distortions \cite{Njobuenwu2015}, when the droplet size is above the characteristic dissipative scale, or laminar distortions \cite{Biferale2014,Spandan2016}, for smaller droplets. It has to be noted that analytical models cannot be used to describe the deformations of large droplets accurately and in particular MM fails to capture non-ellipsoidal deformations. Therefore, it is crucial to develop ab initio models, such as multicomponent LBM with appropriate boundary schemes in order to enforce time dependent fluid deformations. If combined with a Lagrangian history of a turbulent strain matrix, the model allows for a comprehensive characterization of the statistics of droplet shape, size and orientation in a realistic turbulent environment \cite{Biferale2014}. A key parameter to quantify the reaction of the droplet to the time-dependent signal is the ratio between the droplet relaxation time $t_{\tiny \mbox{d}}=\eta_{\tiny \mbox{d}} R / \sigma$ ($R$ being the droplet radius at rest and $\sigma$ the surface tension at the non-ideal interface) and the fluid time scale $t_{\tiny \mbox{f}}=R / u_0$ ($u_0$ is the maximal shear flow intensity). Depending on the ratio $t_{\tiny \mbox{d}}/t_{\tiny \mbox{f}}$ the droplet is either ``enslaved'' ($t_{\tiny \mbox{d}}/t_f \rightarrow 0$) to the fluid variations, or starts to decouple when $t_{\tiny \mbox{d}}/t_{\tiny \mbox{f}} \approx 1$; this influences its deformation and possibly the allignment with the flow. Furthermore, a turbulent signal has a broad spectrum rather than a single time scale $t_{\tiny \mbox{f}}$, thus resulting in a multi-chromatic behaviour coupled to the non-linear response of the droplet deformation process. This is an ideal workspace for LBM mesoscopic models to operate: they intrinsically allow for both droplet deformation at the mesoscale, and they can be constructed to reproduce the desired hydrodynamical flow at large scales. Indeed, droplet deformation properties have been the subject of various papers \cite{Minale08,Greco02,Guido11,Minale04,Minale10,Yue05,VanDerSman08,Komrakova14,Xi99,Sibillo06,Chaffey}, but these typically contain studies of deformation and orientation in steady state flows \cite{Onishi1,VanDerSman08,Komrakova14,Xi99}, or studies of the critical droplet break up condition \cite{GuptaSbragagliaScagliarini2015,RenardyCristini01}, with particular emphasis on the comparison between the (diffuse interface) hydrodynamics of LBM and the sharp interface results \cite{GuptaSbragaglia2014,GuptaSbragagliaScagliarini2015,VanDerSman08,Komrakova14}. Droplet dynamics has also been simulated \cite{GuidoRev,Sibilloetal06,MaffettoneGreco04,MaffettoneGrecoSimeoneGuido,Stone,Verhulst07,Vananroye08}, but the associated quantitative validation has been scarcely detailed in the literature. Our paper aims at filling this gap from the methodological point of view: after revisiting the validation of LBM for steady state flows, we will switch-on time dynamics in controlled situations and quantitatively compare LBM against the analytical predictions of ellipsoidal models at changing the ratio between the droplet relaxation time  $t_{\tiny \mbox{d}}$ and the fluid time scale $t_{\tiny \mbox{f}}$. The paper is organized as follows: in sect.~\ref{sec:problem} we will outline the problem of time dependent droplet deformation in the applicability regime of the MM equation. Section~\ref{sec:static} gives a brief overview on the static deformation Lattice Boltzmann simulations benchmarked against relevant theoretical models. In sect.~\ref{sec:single_phase} the behaviour of a simple oscillatory shear in a 2D LBM channel flow is tested, which will be relevant for sect.~\ref{sec:droplet_time} where we investigate the response of an isolated droplet to a time dependent oscillatory channel flow in a 3D LBM model.

\section{Problem Statement and Continuum Equations}
\label{sec:problem}

We primarily focus on the morphology of droplets in time-dependent laminar flows, where the droplet is simulated via a Lattice Boltzmann algorithm. In the first part of this paper we consider the deformation of a single droplet via a linear shear flow (both static and time-dependent).  We also investigate the deformation (almost) exclusively in
the linear flow regime (i.e. spatially constant shear rate), so that we can compare our simulation results with a phenomenological model, the MM model \cite{MaffettoneMinale98}. In the MM model the droplet is always ellipsoidal, so that we can describe it via a second rank tensor $M_{ij}$, also referred to as the morphology tensor. Droplet deformations are characterised via the components of $M_{ij}$ (e.g. for an undeformed droplet $M_{ij} = \delta_{ij}$). The time evolution of $M_{ij}$ due to an external flow field is given by the MM equation:

\begin{align}
\label{eq:mm_general}
\frac{d M_{ij}}{dt} & = \mbox{Ca} \left [ f_2 (S_{ik} M_{kj} + M_{ik} S_{kj}) + \Omega_{ik} M_{kj} - M_{ik} \Omega_{kj} \right ] \notag \\
& - f_1 \left ( M_{ij} - 3 \frac{III_M}{II_M} \delta_{ij} \right ),
\end{align}

where $S_{ij}$ and $\Omega_{ij}$ are the symmetric and anti-symmetric parts of the shear tensor. $III_M = \det (M_{ij})$ and $II_M \equiv \frac{1}{2} (M_{kk}^2 - M_{ij} M_{ij})$ are the third and second tensor invariants of $M_{ij}$. The capillary number which serves as a control parameter for the droplet deformation is given by the ratio of viscous and interfacial forces of the droplet

\begin{equation}
\label{eq:capillary}
\mbox{Ca}=\frac{\eta_{\text{s}} R G}{\sigma},
\end{equation}

where $G$ is the shear rate of the surrounding flow. Unlike previous analytical approaches to model droplet deformation in laminar flows \cite{Taylor32} the MM model is not based on a perturbative expansion in the capillary number $\mbox{Ca}$. Thus we can increase $\mbox{Ca}$ to relatively large values in our LBM simulations, since we have a robust analytical model to compare it with. However, it should be noted, that the MM model requires the droplet shape to be ellipsoidal at all times (an ad hoc assumption). The deformation is thus defined as $D \equiv \frac{L - W}{L + W}$, where $L$ and $W$ are the major and minor ellipsoidal axes respectively. Since we want to compare the LBM simulations with the MM-model, we need to make sure that we remain in the linear flow regime and check that our deformed LBM droplet is actually ellipsoidal at all times. Since we investigate droplets in confined systems in particular, we remark that the MM model has to be modified to account for a confined droplet. This can be achieved by modifying the parameters $f_1$ and $f_2$ in eq.~(\ref{eq:mm_general}) for the confined case. In the unbounded case \cite{MaffettoneMinale98}, which we call MM-unbounded, we have

\begin{align}
\label{eq:f_unbounded}
f_1^{\text{un}}(\chi) & = \frac{40 (\chi + 1)}{(3 + 2 \chi) (16 + 19 \chi)}, \notag \\
f_2^{\text{un}}(\chi, \mbox{Ca}) & = \frac{5}{3 + 2 \chi} + \frac{3 \mbox{Ca}^2}{2 + 6 \mbox{Ca}^2},
\end{align}

and for the confined case \cite{Minale08} which we call MM-confined

\begin{align}
\label{eq:f_confined}
f_1 (\chi, \alpha) & = \frac{f_1^{\text{un}}(\chi)}{1 + f_1^c(\chi) \mbox{C}_\text{s} \frac{\alpha^3}{8}}, \notag \\
f_2 (\chi, \mbox{Ca}, \alpha) & = f_2^{\text{un}}(\chi, \mbox{Ca}) \left (1 + f_2^c(\chi) \mbox{C}_\text{s} \frac{\alpha^3}{8}, \right )
\end{align}

with 

\begin{align}
\label{eq:f_critical}
f_1^c(\chi) & = \frac{44 + 64 \chi - 13 \chi^2}{2 (1 + \chi) (12 + \chi)}, \notag \\
f_2^c(\chi) & = \frac{9 \chi - 10}{12 + \chi},
\end{align}

$\mbox{C}_\text{s}$ denotes a form factor depending on the degree of confinement \cite{ShapiraHaber90} and $\alpha \equiv \frac{2 R}{L_z}$ is  the aspect ratio of the droplet length scale to the scale of the confinement (e.g. the width of a channel). The form factor is chosen according to \cite{Guido11} as $\mbox{C}_\text{s} = 5.6996$ throughout our simulations, since the droplet's centre of mass is located in the middle between two channel walls. Moreover the viscous ratio $\chi \equiv 1$ and the density ratio $\rho_{\tiny \mbox{d}} / \rho_{\tiny \mbox{s}} \equiv 1$ in all preceding calculations.

\section{Multicomponent Lattice Boltzmann scheme and boundary conditions}
\label{sec:scmc_bc}

\begin{figure}[!htbp]
\centering
\includegraphics[scale=0.65, trim={10mm, 125mm, 20mm, 75mm}, clip]{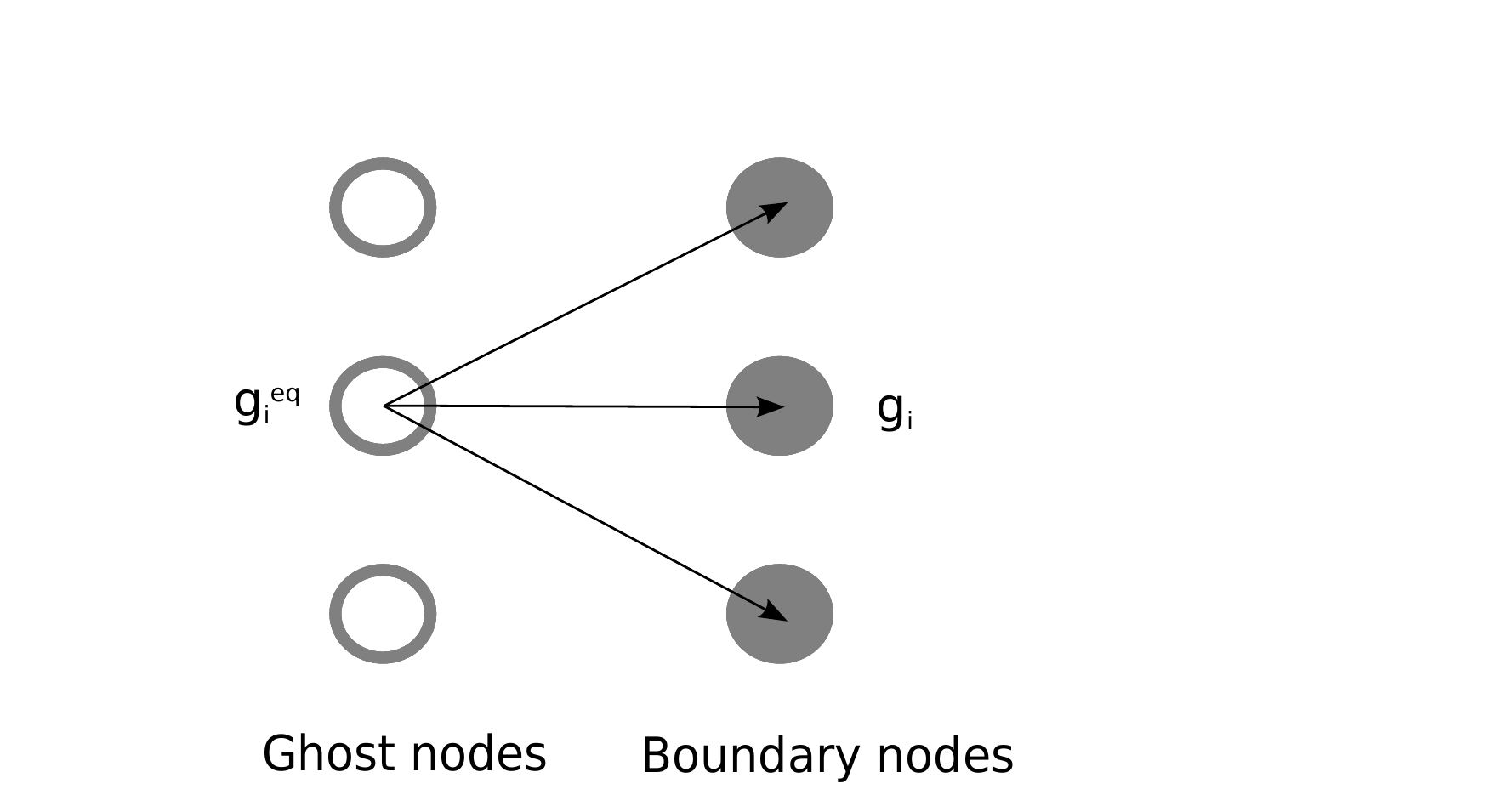}
\caption{Sketch of the streaming step from ghost to boundary nodes. The ghost nodes are initialised via the local equilibrium distributions $g_i^{\text{eq}}$. By initialising the ghost nodes with a given equilibrium density $\rho(\vec{x},t)$ and equilibrium velocity field $\vec{u}(\vec{x},t)$ we can effectively set the boundary conditions of the system.}
\label{fig:halos}
\end{figure}

The classical Lattice Boltzmann Model (LBM) for single phase flows needs to be modified to account for a system containing two immiscible fluids, in particular the fluid-fluid interface between them. One of the most used scheme to model the fluid-fluid interface is the Shan-Chen Multi-Component model (SCMC) \cite{Shan93,Shan94}. For two (or more) immiscible fluids we need to distinguish between the type of fluid component at hand, thus we get for the mass and momentum densities:

\begin{align}
\label{eq:multi_momentum}
\rho(\vec{x}, t) & = \sum_{\sigma} \sum_i g_i^{\sigma}(\vec{x}, t), \notag \\
\rho(\vec{x}, t) \vec{u}(\vec{x}, t) & = \sum_{\sigma} \sum_i g_i^{\sigma}(\vec{x}, t) \vec{c}_i,
\end{align}

where $g_i^{\sigma}(\vec{x},t)$ denotes the populations in the LBM model for the fluid component $\sigma$ and $\vec{c}_i$ are the lattice velocities. The interaction at the respective fluid-fluid interface \cite{Sbragaglia2013,Sega2013} is given by:

\begin{equation}
\label{eq:scmc_interface}
\vec{F}^{\sigma}(\vec{x}) = - \psi_{\sigma}(\vec{x}) \sum_{\sigma' \neq \sigma} \sum_{i=1}^{N} \mathcal{G}_{\sigma, \sigma'} w_i \psi_{\sigma'}(\vec{x} + \vec{c}_i) \vec{c}_i,
\end{equation}

where $\psi_{\sigma}(\vec{x})$ is a local pseudo-potential which may be defined via the phase densities $\rho_{\sigma}(\vec{x},t)$. $\mathcal{G}_{\sigma,\sigma'}$ is a coupling constant for the two phases $\sigma$ and $\sigma'$ at position $\vec{x}$ and $w_i$ are the lattice isotropy weights. One should note that the stencil for the SCMC pseudo- potential interaction does not necessarily have to coincide with the stencil populations for the LBM streaming, but could be a different lattice stencil altogether, given that the interaction force $\vec{F}^{\sigma}(\vec{x})$ remains isotropic. In order to effectively simulate a time-dependent flow we shall use specially modified boundary conditions. We make use of the ghost populations (or halos) to store the local LBM equilibrium population distributions given by the systems boundary values for the density $\rho(\vec{x}, t)$ and velocity  $\vec{u}(\vec{x},t)$ of the outer fluid (for simplicity we will treat a single component fluid).

\begin{equation}
\label{eq:equilibrium}
g_i^{\text{eq}} (\vec{x},t) = \rho(\vec{x},t) w_i \left ( 1 + 3 \, \vec{c}_i \cdot \vec{u} + \frac{9}{2} (\vec{c}_i \cdot \vec{u})^2 - \frac{3}{2} \vec{u}^2 \right )
\end{equation}
 
with $\{w_i\}$ being the lattice weights for the set of lattice vectors $\{\vec{c}_i\}$. Thus the ghost distributions will update the boundary nodes during the LBM streaming step and let the system know about the previously chosen boundary conditions (see fig.~\ref{fig:halos}).
The streaming and collision steps are given by the Lattice Boltzmann eq.:

\begin{equation}
\label{eq:lbe}
g_i(\vec{x} + \vec{c}_i \Delta t, t + \Delta t) - g_i(\vec{x}, t) = \Omega(\{g_i(\vec{x},t)\}),
\end{equation}

where $\Omega(\{g_i(\vec{x},t)\})$ is the collision operator depending on the whole (local) set of lattice populations and $\Delta t$ is the simulation time step. For MRT (multi- relaxation time scale) the collision operator is linear and contains several relaxation times linked to its relaxation modes (depending on the lattice stencil) \cite{Humieres02}. One relaxation time $\tau$ is directly linked to the kinematic viscosity $\nu$ in the system

\begin{equation}
\label{eq:viscosity}
\nu = \frac{1}{3} \left ( \tau - \frac{1}{2} \right ),
\end{equation}

which is one of the primary links between the LBM scheme and hydrodynamics \cite{Benzi92,Succi01}. Since the ghost nodes only stream into the system and not out of it, we have to correct the local population mass densities in order to keep the system mass conserving \cite{Mattila09,Hecht10,Zou97}. Even though the general idea of this boundary scheme is particularly useful for unbounded systems, we willl use it in the confined case for our simulations as well. Thus, in our case the boundary scheme is equivalent to a mid-bounce-back rule. However, it should be noted that this scheme may be extended to account for pressure driven boundary conditions in unbounded systems \cite{Mattila09,Hecht10}.

\begin{figure}[!htbp]
\centering
\includegraphics[scale=0.65]{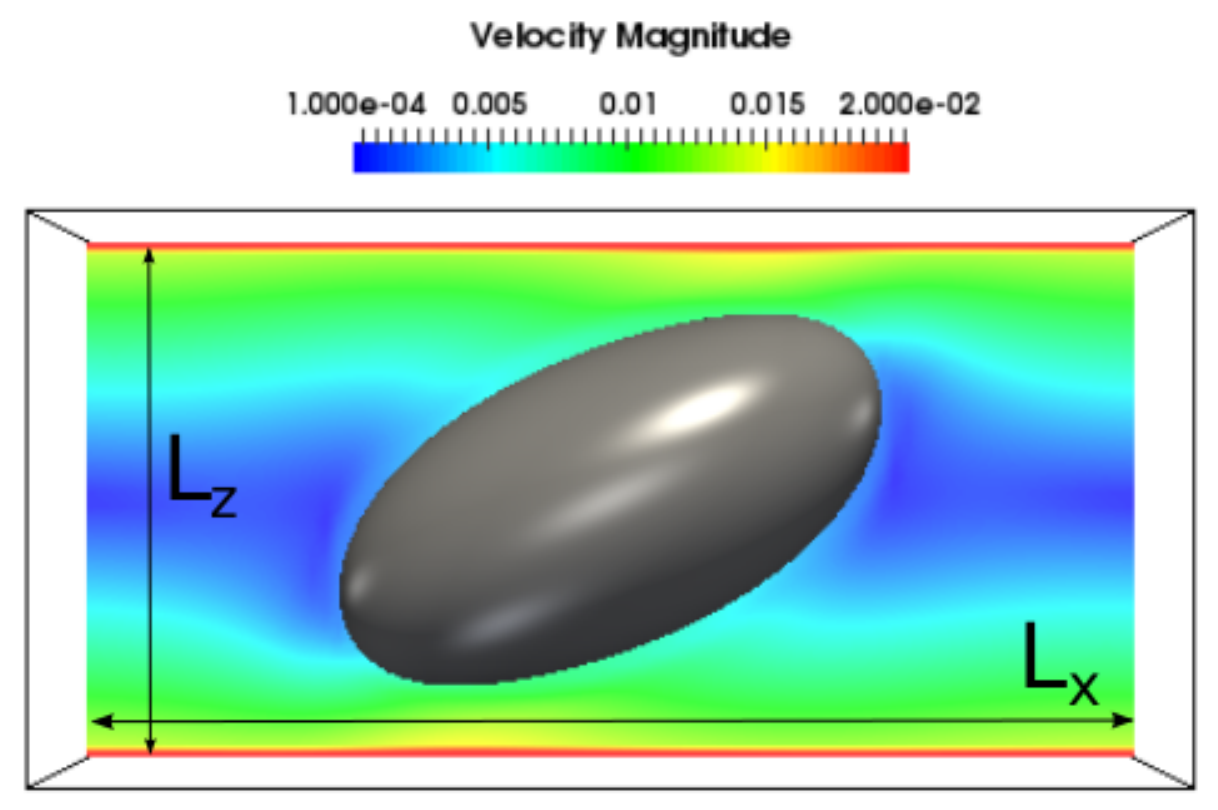}
\caption{Screenshot of a multicomponent LBM simulation. A droplet is ellipsoidally deformed via an external shear flow, created by the moving channel walls. The relevant system parameters are: the initial droplet radius $R$, the shear rate $G$, the channel width $L_z$ and the other lengths of the simulation domain $L_x$ and $L_y$ (not shown). The magnitude of the overall velocity field in lbu is given via a colour gradient.}
\label{fig:sketch_droplet}
\end{figure}

\begin{figure}[!htbp]
\centering
\textbf{Lower Resolution}
\includegraphics[scale=0.65]{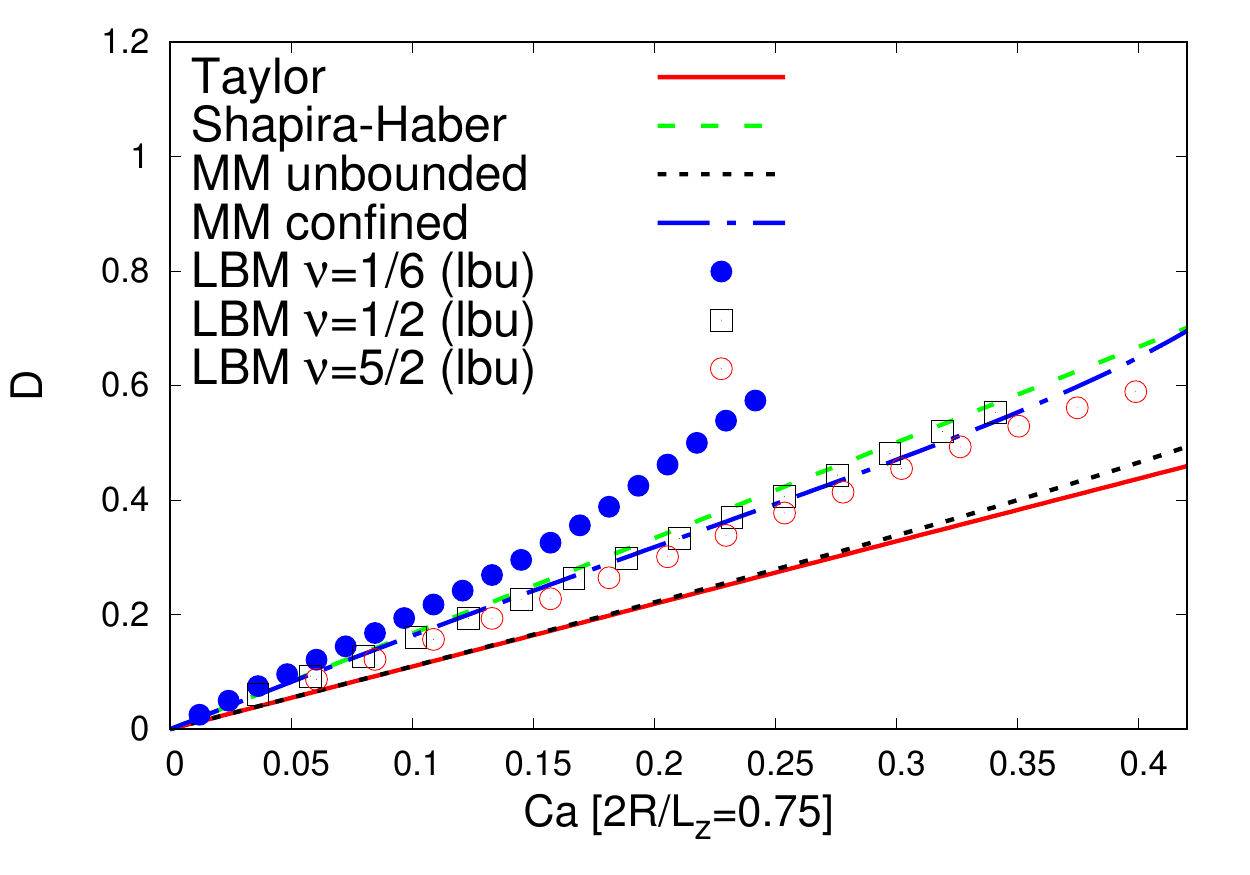}
\textbf{Higher Resolution}
\includegraphics[scale=0.65]{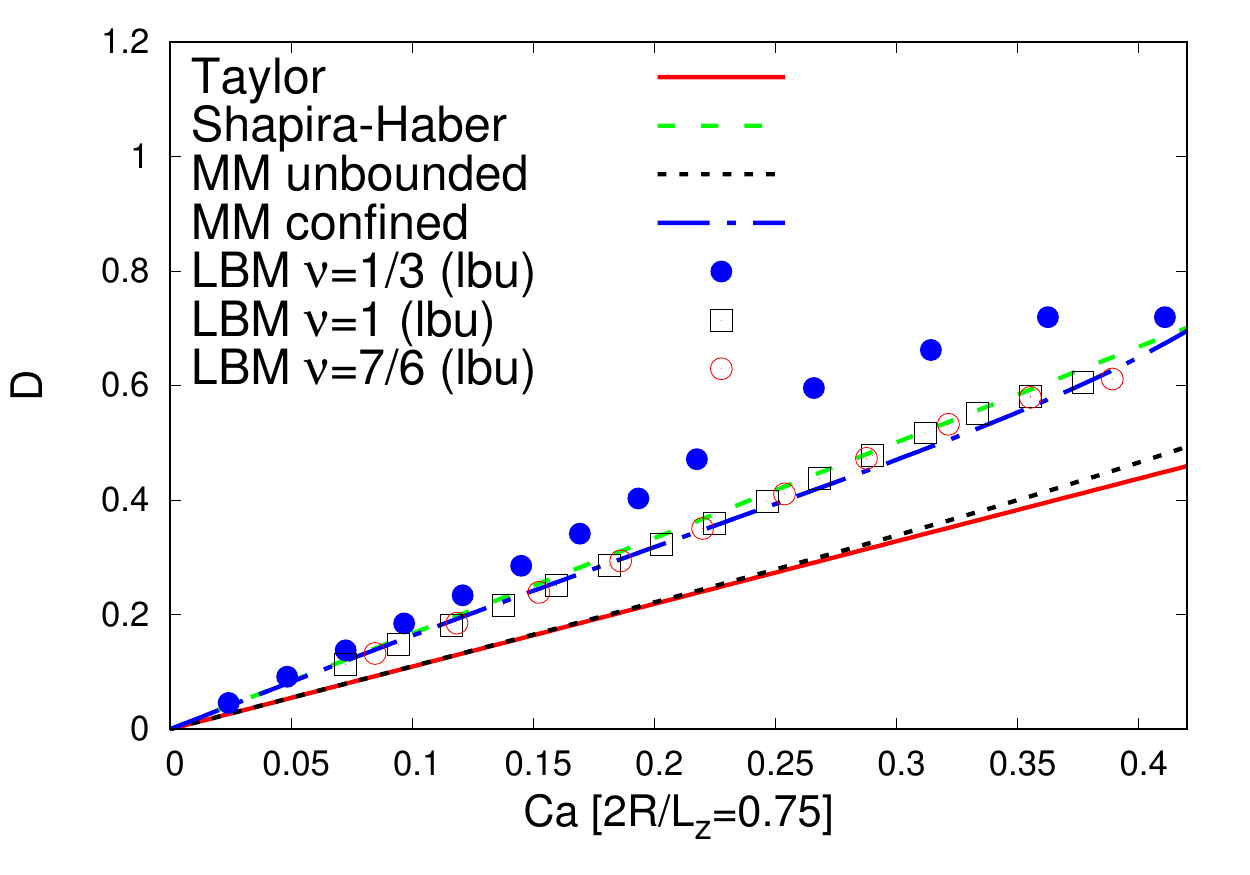}
\caption{Droplet deformation test  benchmarked against several theoretical models: perturbative models in the capillary number $\mbox{Ca}$, Taylor  (unconfined droplet) \cite{Taylor32} and Shapira-Haber (accounting for droplet confinement) \cite{ShapiraHaber90} and models extending the $\mbox{Ca}$ range to higher values, MM-unbounded \cite{MaffettoneMinale98} and the MM-confined \cite{Minale08} model.  The extent of the deformation measured by the parameter $D$ is plotted against the capillary number $\mbox{Ca}$.  We choose a series of three Reynolds number ranges by selecting three kinematic viscosities. We can see that the agreement between the LBM simulation and the theoretical predictions improves significantly for lower Reynolds numbers. This is due to the reduction of inertia in the system. Top panel: Resolution $80 \times 40 \times 40$ Bottom panel: Resolution $160 \times 80 \times 80$. To account for similar Reynolds number ranges at a higher resolution the simulations in the bottom panel have higher viscosity values than the one in the top panel.}
\label{fig:droplet_static}
\end{figure}

\section{Static droplet deformation}
\label{sec:static}

The first step is to test our algorithm in the case of the deformation of a single droplet in a constant shear flow confined in a channel, see fig.~\ref{fig:sketch_droplet}. Inertia is characterised by the Reynolds number $\mbox{Re} = R^2 G / \nu$ where, in the case of a simple shear flow, the shear rate is $G = 2 u_0 / L_z$, with $u_0$ being the maxmimum shear at the wall and $L_z$ the channel width. Now we let the droplet evolve in the shear flow and measure its deformation. We consider only set-ups with an aspect ratio $\alpha \equiv 2 R / L_z = 0.75$ and keep the viscosity ratio $\chi = 1$ throughout all simulations. According to \cite{Ioannou16} the droplet will be stable up to a value of $\mbox{Ca} \approx 0.4$ regardless of our choice for the confinement ratio $\alpha$. A series of LBM runs is shown in fig.~\ref{fig:droplet_static} and three different values for the kinematic viscosity $\nu$ in lbu (Lattice Boltzmann units). We may see that for the lowest value of $\nu$ the deformation $D$ is deviating substantially from the theoretical predictions for a confined droplet, given both by the Shapira-Haber model \cite{Guido11} and the MM-confined model \cite{Minale08}. 
For the two lower $\mbox{Re}$ values the simulations agree much better with the MM-confined model predictions. Figure~\ref{fig:droplet_static} also shows the static deformations for a higher resolution in the bottom panel where the respective Reynolds number $\mbox{Re}$ lies in the same range as for the plots in the top panel. In this case the droplet deformation of our LBM scheme agrees even better with the theoretical predictions for a confined droplet. Thus we can deduce both: that we need a significantly low Reynolds number and that we may only compare our simulation results to models which account for the confinement of the droplet. To our knowledge this is the first benchmark of LBM against theoretical predictions for the influence of droplet inertia in static droplet deformation in a system with a significant confinement ratio.

\section{Probing the parameter space: single component oscillating shear flow}
\label{sec:single_phase}

After having benchmarked the static droplet deformation against a variety of theoretical models we investigate the 

\begin{figure}[!htbp]
\centering
\includegraphics[scale=0.5,trim={10mm, 125mm, 20mm, 50mm},clip]{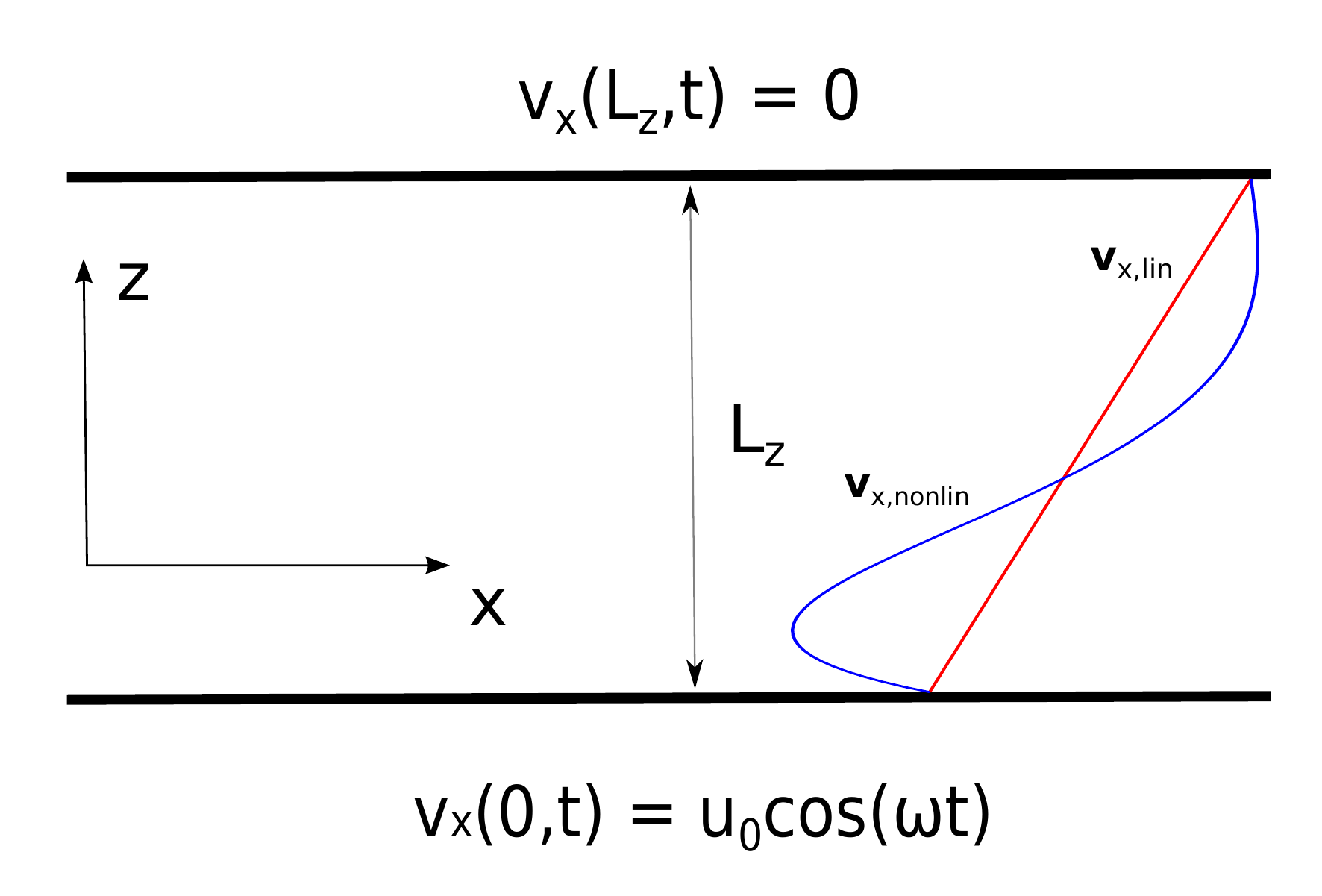}
\caption{Sketch of the single phase model set-up. A 2D channel of width $L_z$ with one stationary and one oscillating wall. The flow in the $x$-direction is periodically extended. Two typical velocity profiles of the oscillating channel flow are shown, a linear one in red and a nonlinear one in blue.}
\label{fig:sketch_channel}
\end{figure}

droplet behaviour under a time-dependent linear shear flow. Before considering explicitly the case of a binary fluid (see sect.~\ref{sec:droplet_time}) we need to determine a suitable range for our LBM parameters. We {\it remark} that LBM works well as a ``hydrodynamic solver'' only if the LBM populations are close to the hydrodynamical manifold. Hence it is crucial to design a set of ``working parameters'' for which we know that our LBM scheme correctly solves the time-dependent hydrodynamical equations. Specifically, for the case of a time-dependent shear flow, we will compare our  LBM scheme against the exact time dependent solution of an oscillating shear flow \cite{Landau}. For simplicity we modify the boundary conditions for a 2D channel flow by setting $v_x(0, t) = u_0 \cos (\omega t)$ and $v_x(L_z, t) = 0$, i.e. one side of the channel is oscillating with a shear frequency $\omega_f = \omega /( 2 \pi)$ and the other one is static (see fig.~\ref{fig:sketch_channel}). Making use of the incompressibility condition $\nabla \cdot \vec{v} = 0$ we obtain for the Navier-Stokes equation:

\begin{equation}
\label{eq:diffusionl}
\pdev{t} \, v_x = \nu \, \pdev{z}^2 \, v_x.
\end{equation}
Making the ansatz
\begin{equation}
\label{eq:ansatz}
v_x(z, t) = e^{-i \omega t} \left ( A \, \cos(k z) + B \, \sin(k z) \right )
\end{equation}
leads to the dispersion relation
\begin{equation}
\label{eq:dispersion}
k = \frac{1 + i}{\delta},
\end{equation}
where $\delta \equiv \sqrt{\frac{2 \, \nu}{\omega}}$ is the penetration depth of the system. The solution for $v_x$ reads
\begin{equation}
\label{eq:complex}
v_x(z, t) = u_0 \, e^{-i \omega t} \, \frac{\sin(k (L_z - z))}{\sin(k L_z)},
\end{equation}

\onecolumn
\begin{figure}[!htbp]
\vcenteredhbox{\includegraphics[scale=0.48]{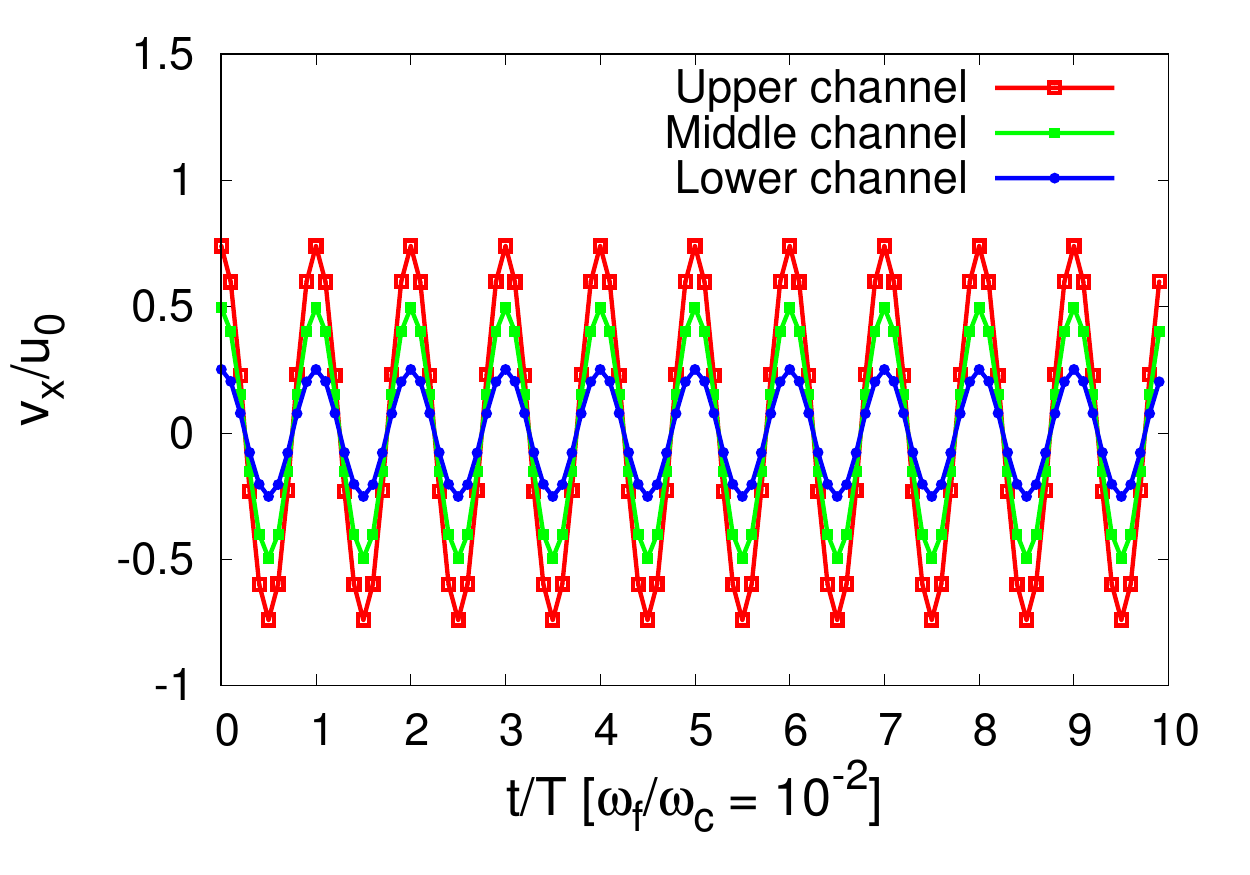}}
\vcenteredhbox{\includegraphics[scale=0.48]{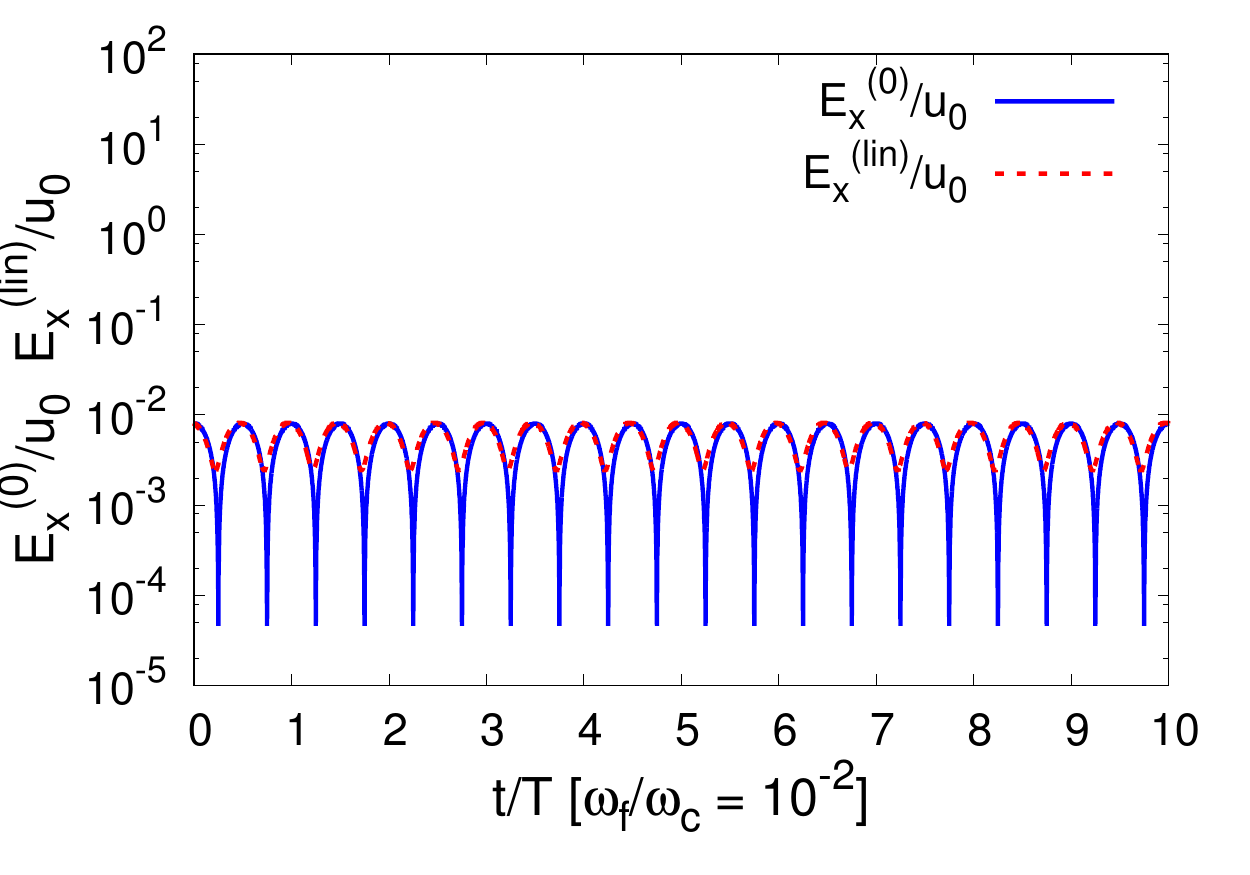}}
\vcenteredhbox{\includegraphics[scale=0.48]{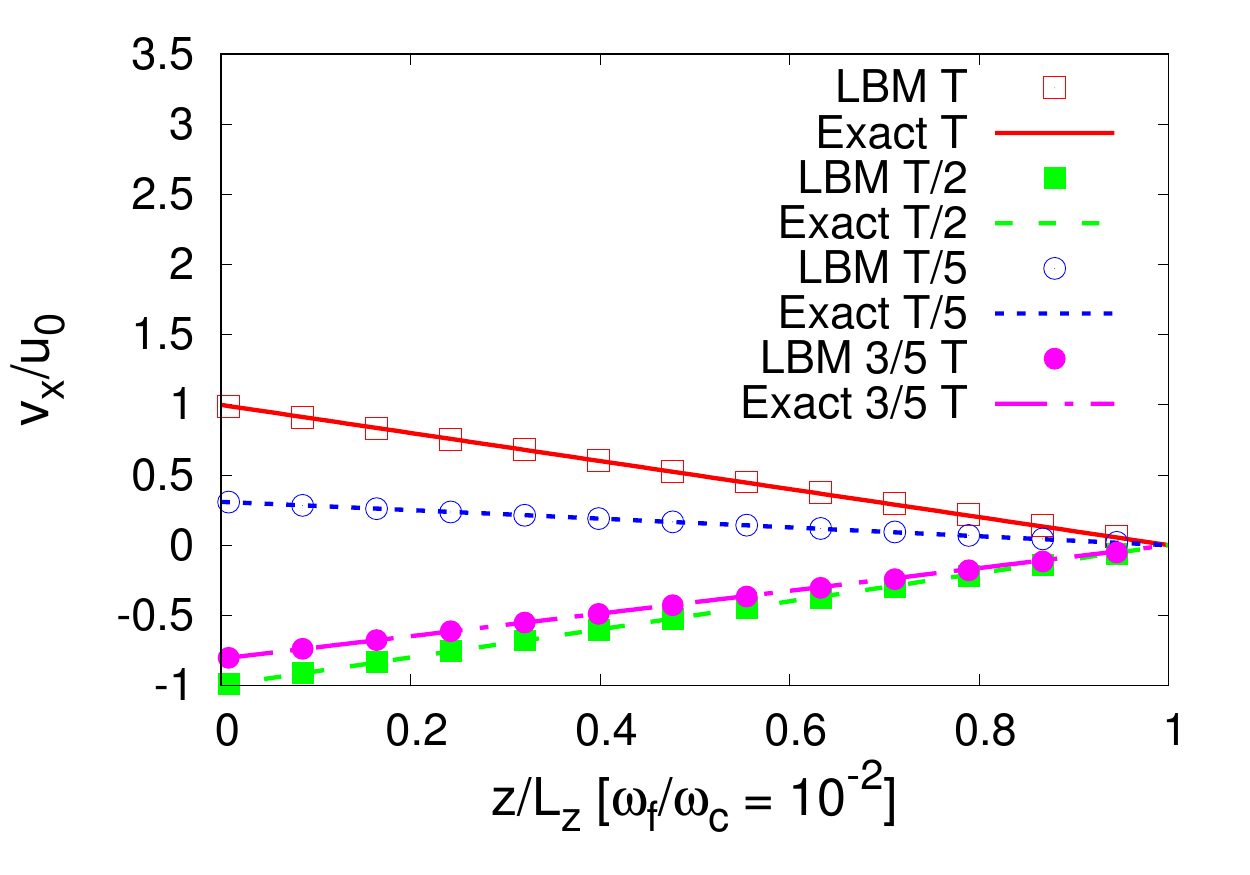}}\\
\vcenteredhbox{\includegraphics[scale=0.48]{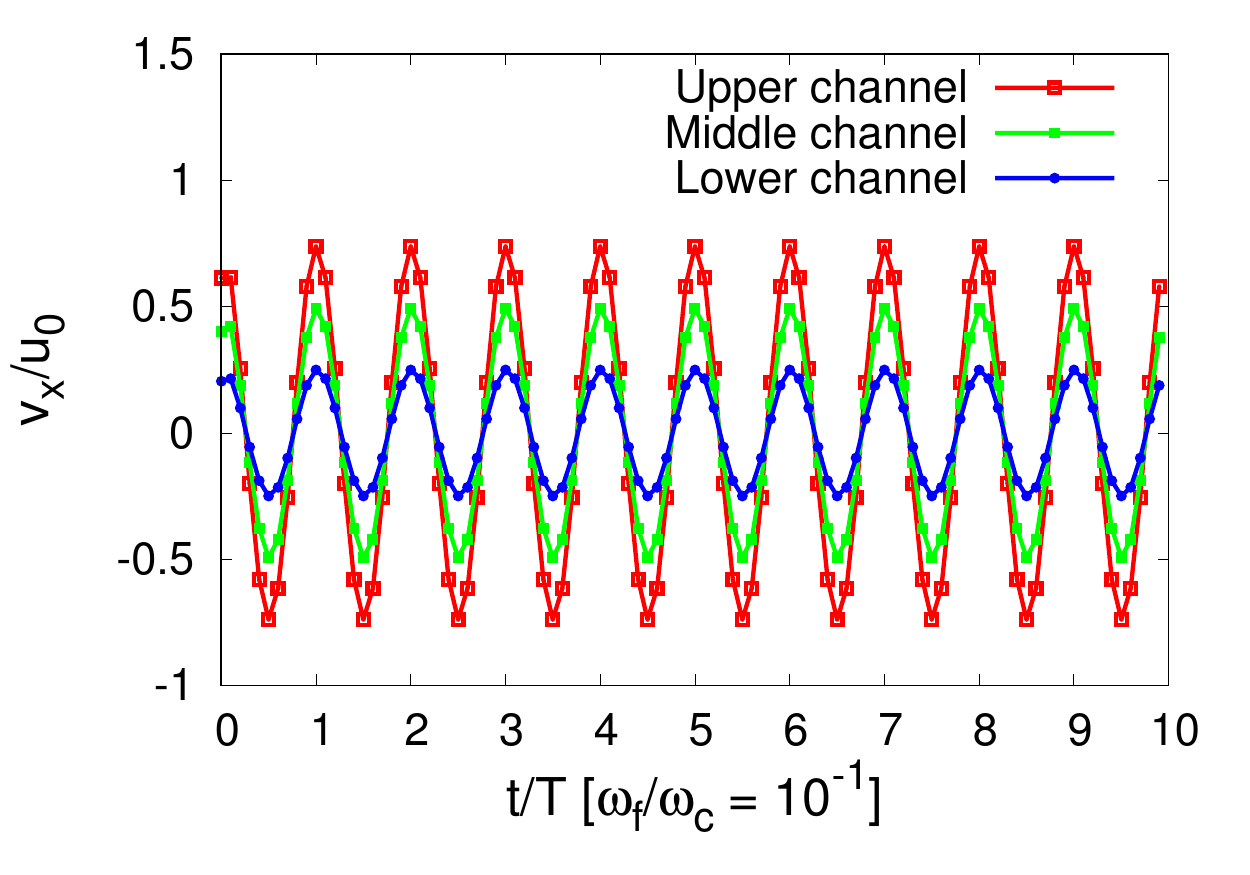}}
\vcenteredhbox{\includegraphics[scale=0.48]{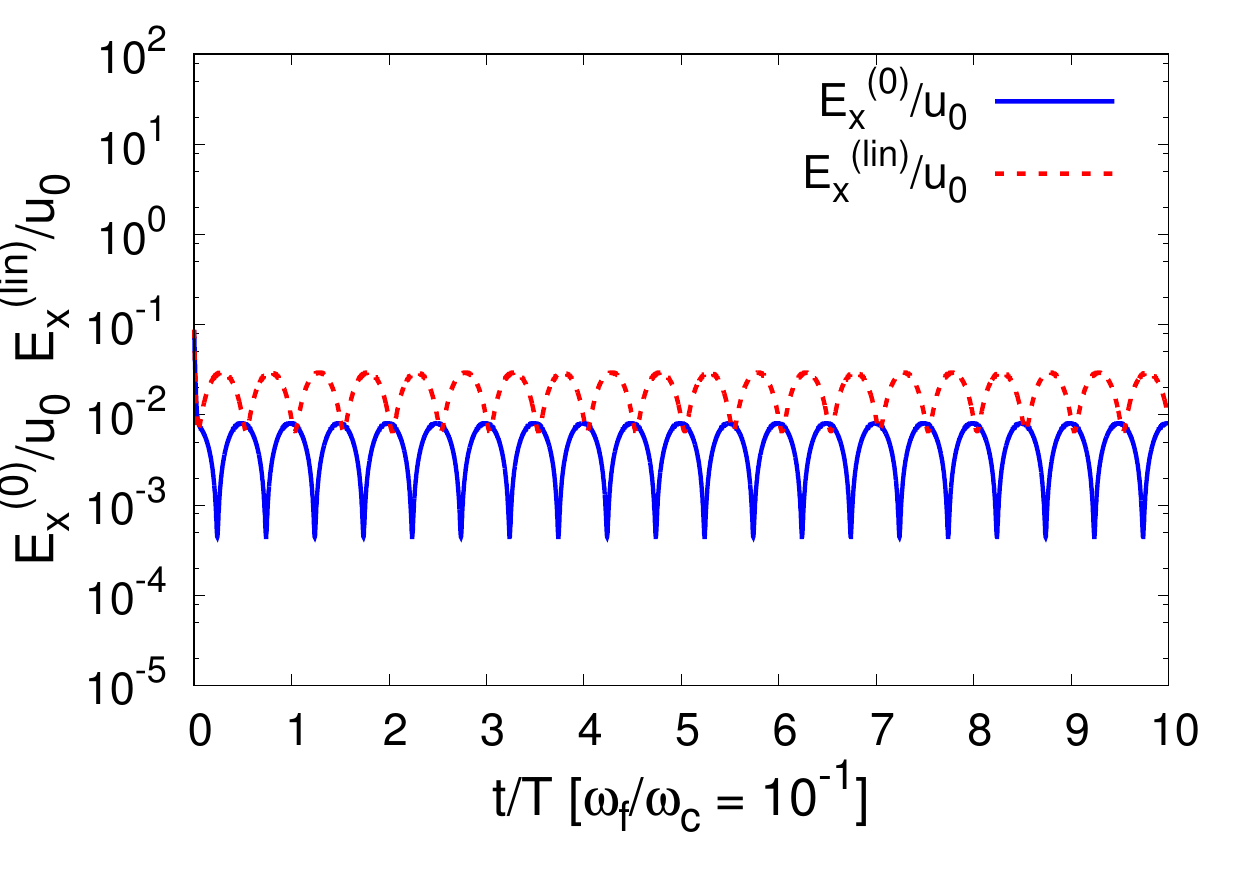}}
\vcenteredhbox{\includegraphics[scale=0.48]{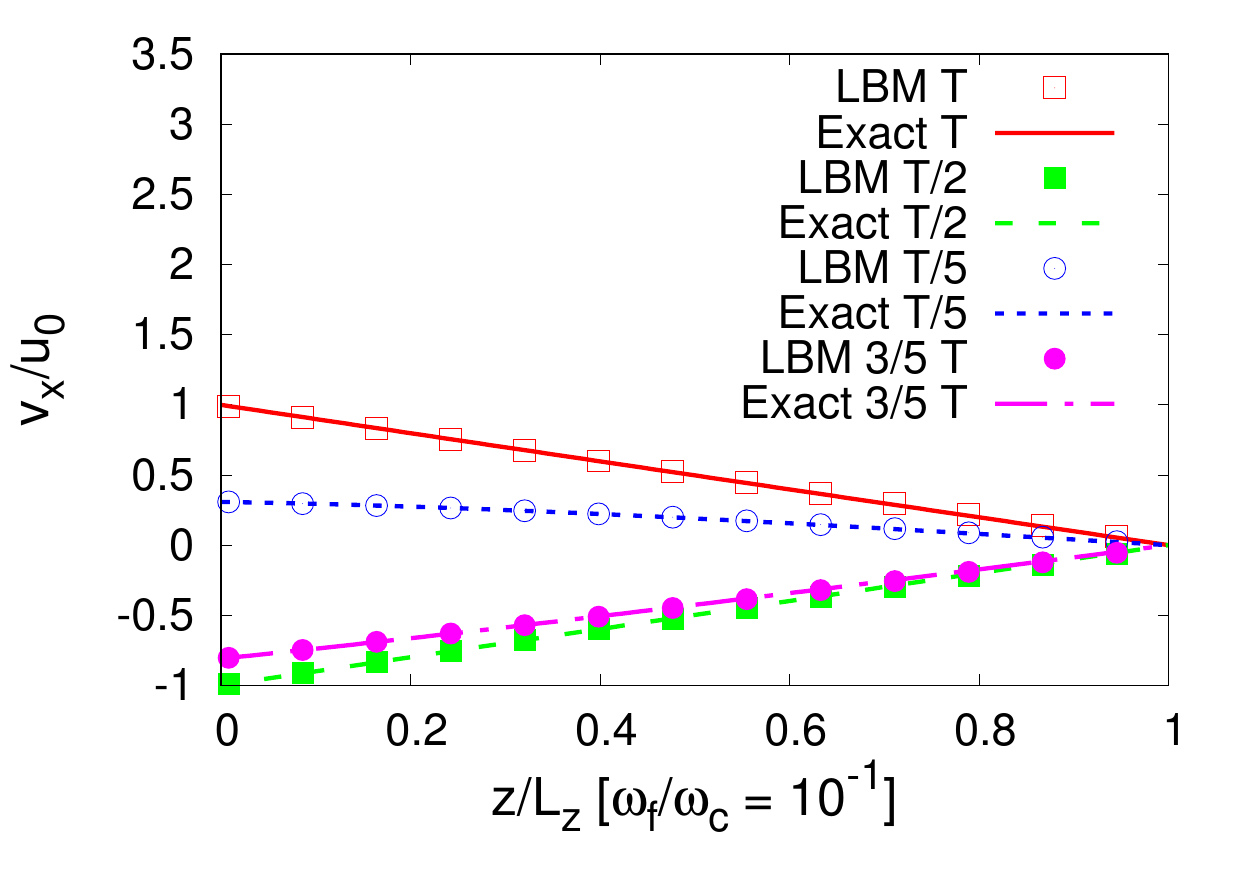}}\\
\vcenteredhbox{\includegraphics[scale=0.48]{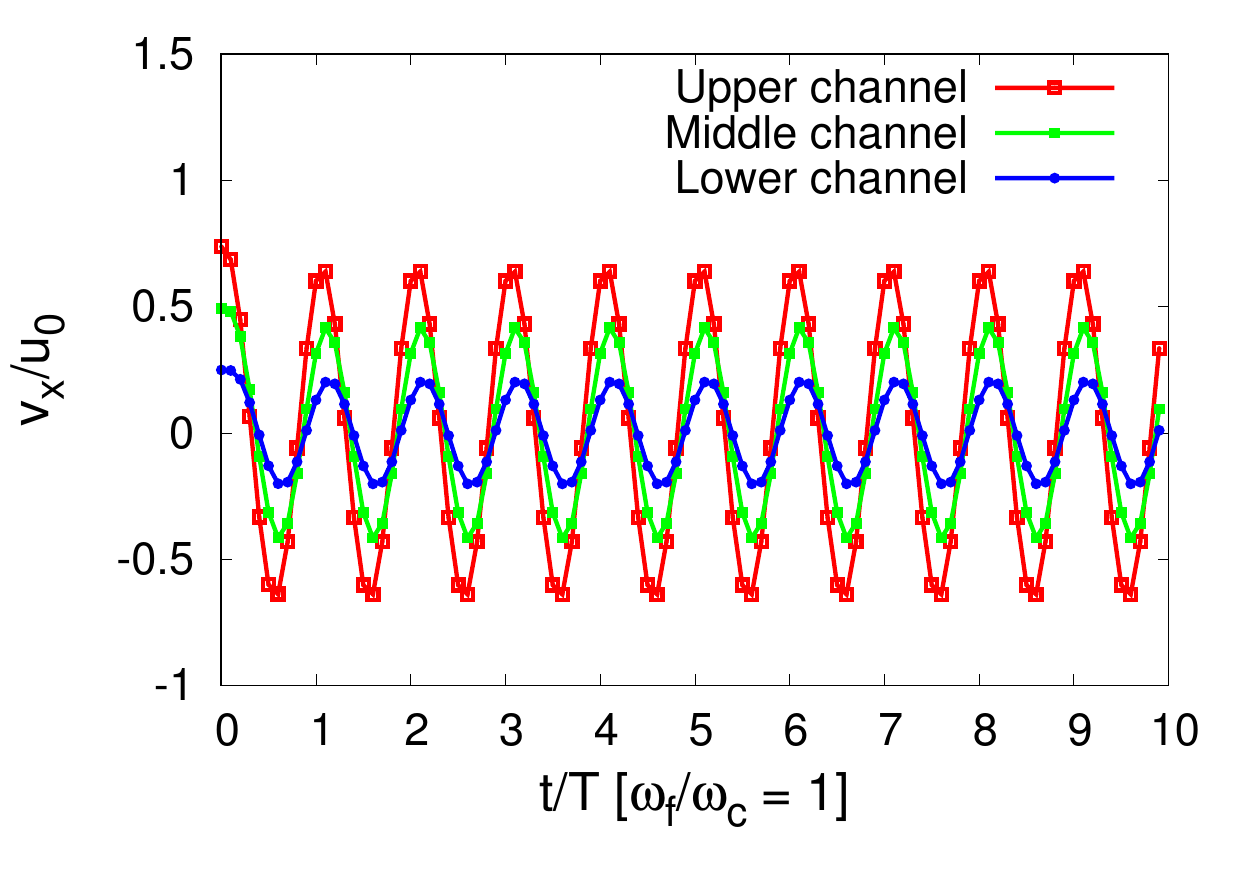}}
\vcenteredhbox{\includegraphics[scale=0.48]{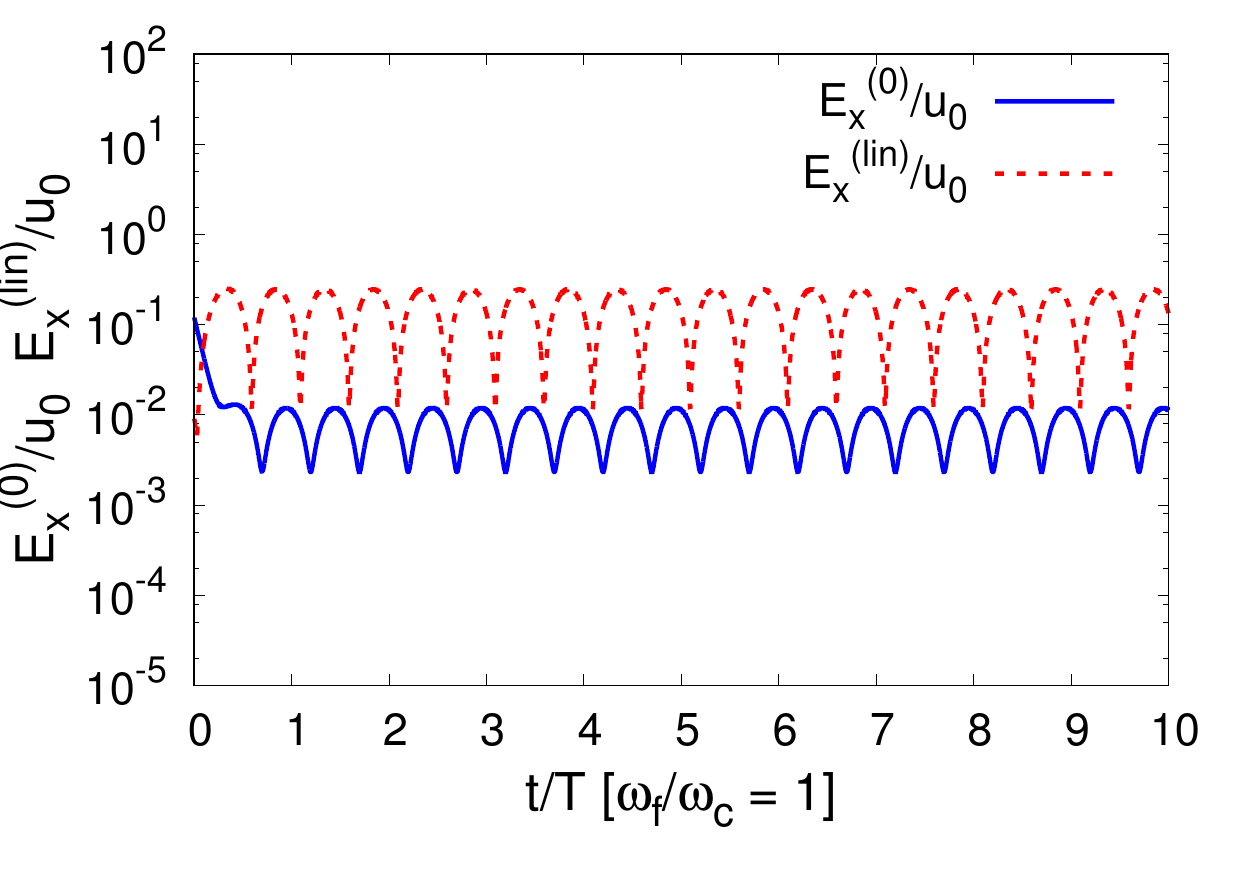}}
\vcenteredhbox{\includegraphics[scale=0.48]{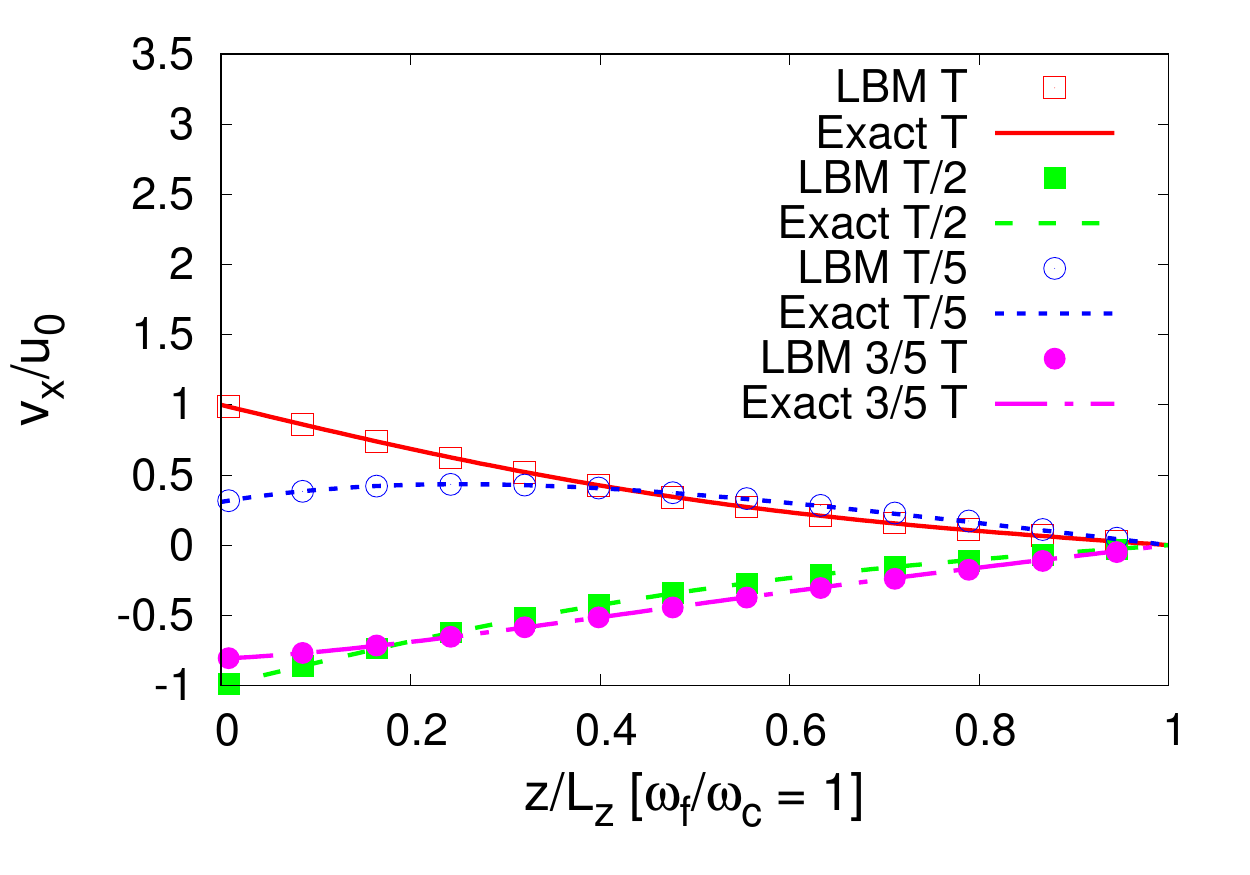}}\\
\vcenteredhbox{\includegraphics[scale=0.48]{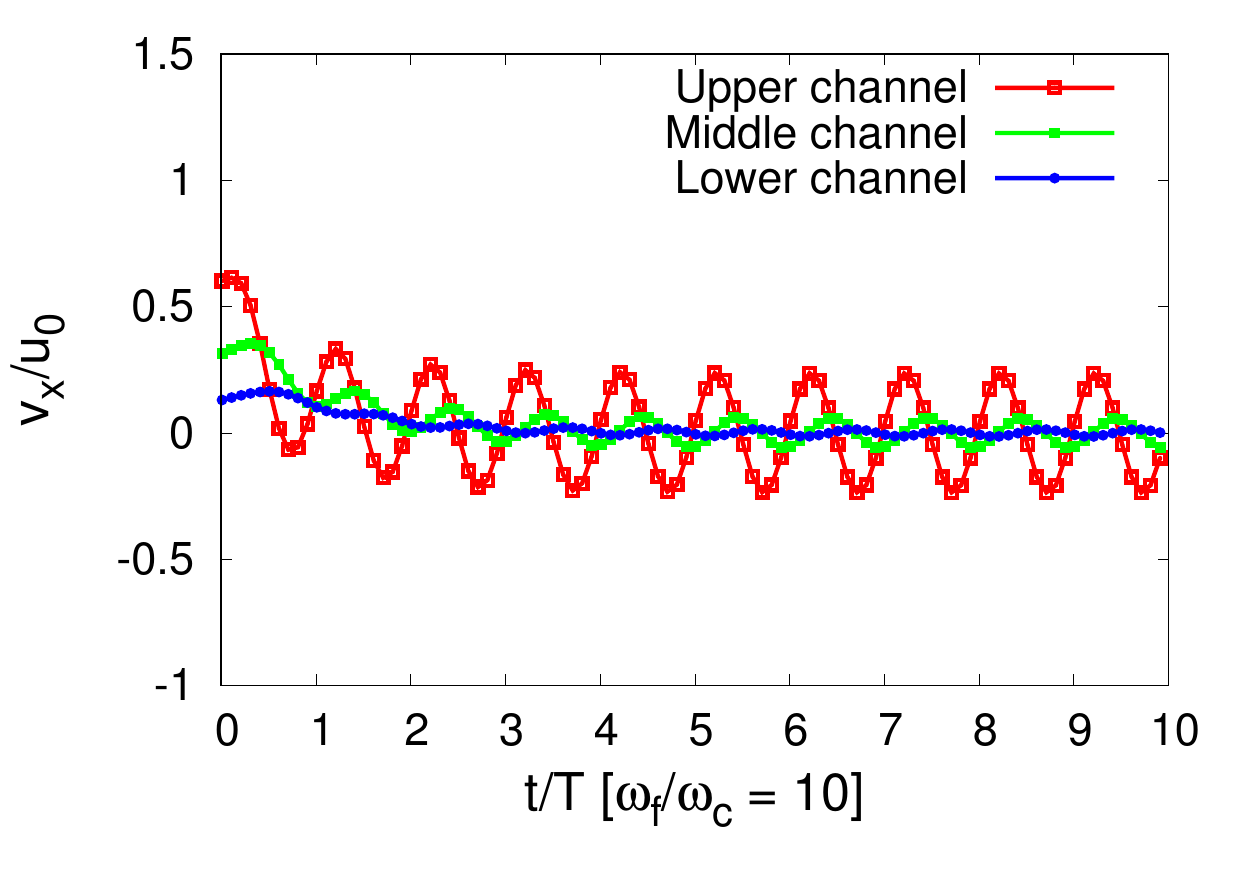}}
\vcenteredhbox{\includegraphics[scale=0.48]{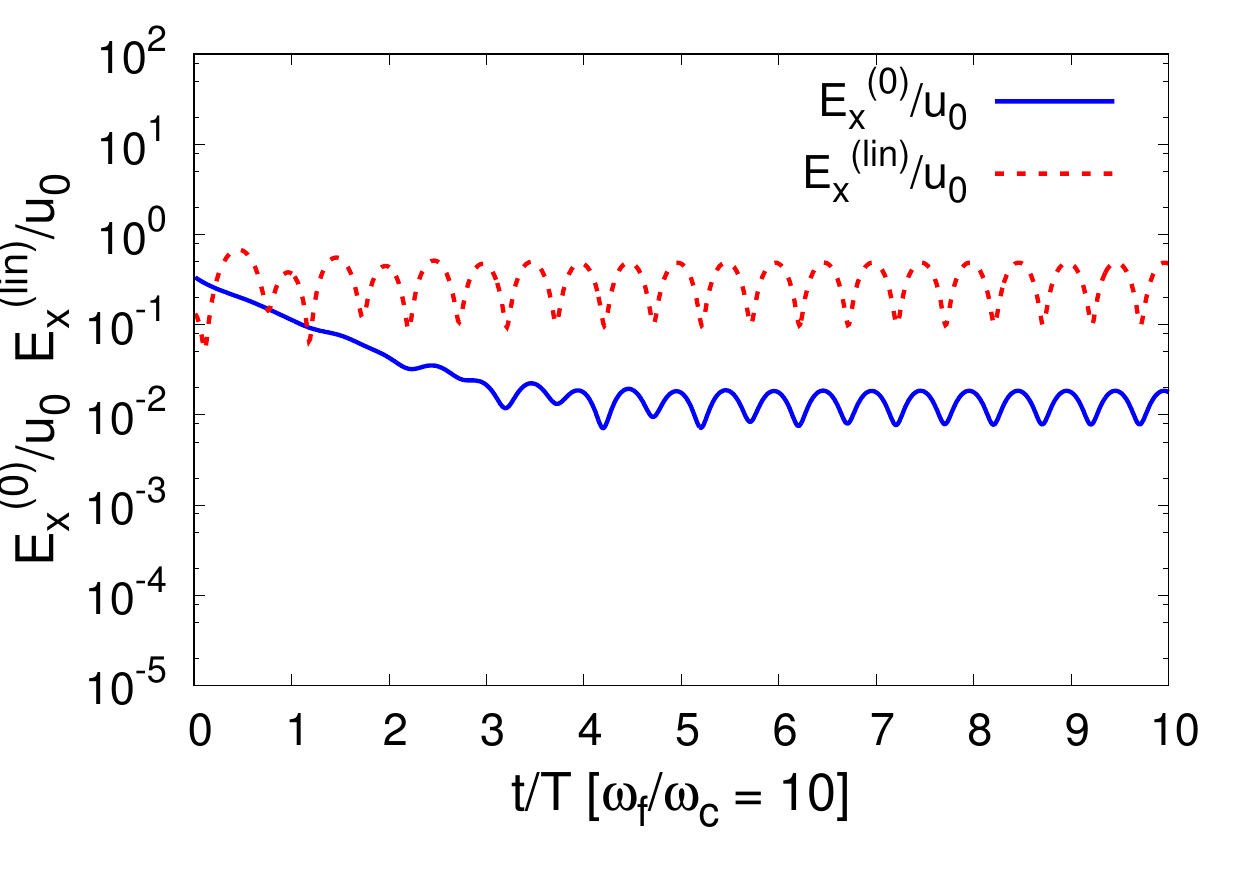}}
\vcenteredhbox{\includegraphics[scale=0.48]{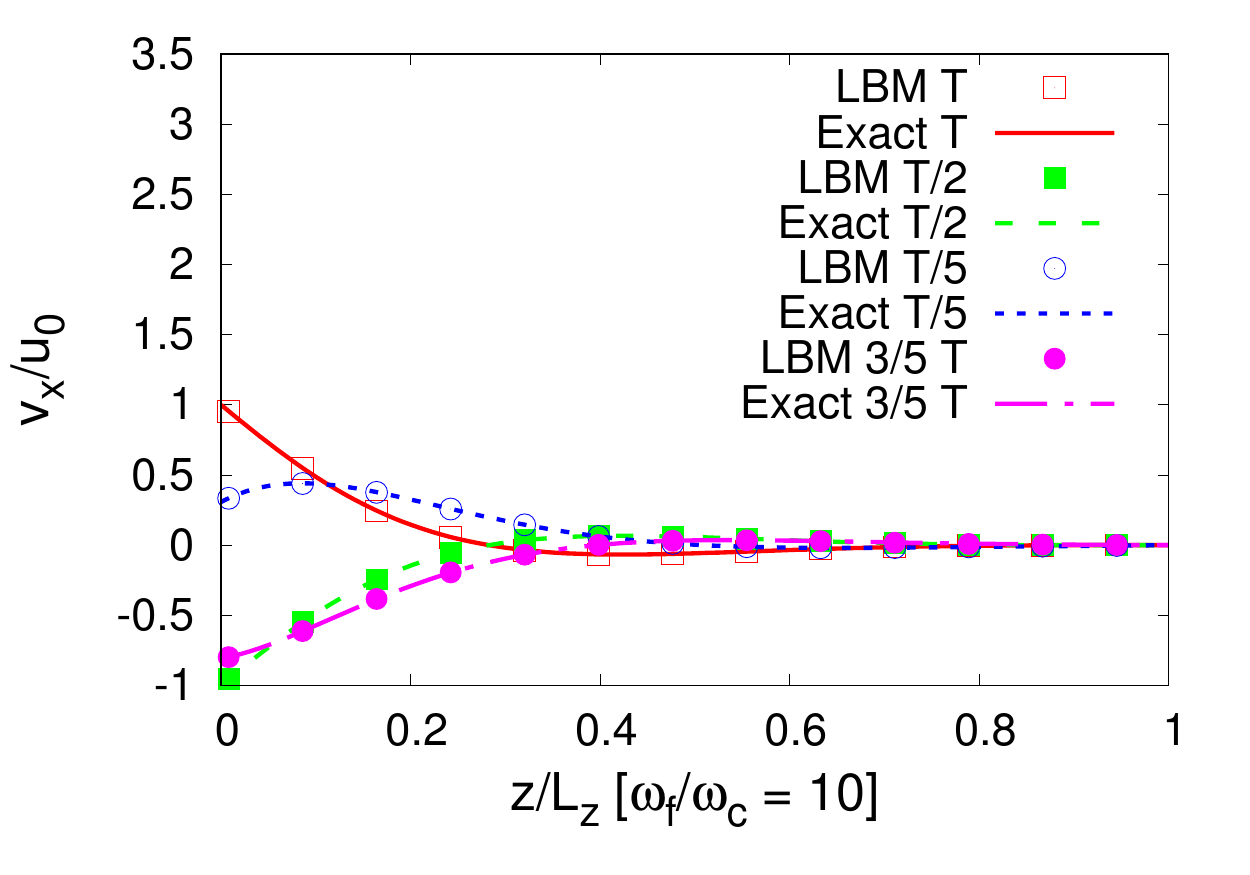}}
\caption{Summary of a LBM benchmark series (channel width $L_z = 128$, relaxation time $\tau=1$, maximal shear velocity $u_0=10^{-3}$) for a single component in the presence of an oscillating shear. Left panels: time evolution of the velocity field at different height locations: upper channel ($z=3 L_z/4$), middle channel ($z=L_z/2$) and lower channel ($z=L_z/4$). We see a transient for higher frequencies $\omega_f / \omega_c \sim 10$, which is due to the relaxation of the velocity field $v(x,z,t)$ to the analytical solution in eq.~(\ref{eq:complex}) from a zero velocity initialisation state. Middle panels: normalized error with respect to the exact solution (eq.~(\ref{eq:error})) and normalized error with respect to the linearised solution (eq.~(\ref{eq:errorlin})). Right panels: Velocity profile $v_x(z,t)$ as a function of the dimensionless cross-flow coordinate $z/L_z$ at different times $t=T/5, T/2, 3 T/5, T$ (in units of the shear period $T = 1 / \omega_f$). In all plots we non-dimensionalise the shear frequency $\omega_f$  by the critical frequency $\omega_c$. We may notice qualitatively that the velocity profile is becoming gradually more nonlinear with increasing frequency $\omega_f$ after passing the critical region $\omega_f/\omega_c \approx 0.1$. This is in agreement with the system parameter scan shown in fig.~\ref{fig:error_collapse}.
\label{fig:time_series}}
\end{figure}
\twocolumn

\twocolumn
\begin{table}[!htbp]
\twocolumn
\caption{System parameter scan of the single phase oscillatory channel flow (see sect.~\ref{sec:single_phase}). A few representative cases are reported of changing the channel width $L_z$, the LBM relaxation time $\tau$, the maximum wall velocity $u_0$, the shear oscillation frequency $\omega_f$.}
\begin{tabular}{ @{} c c c c c c @{} }
\arrayrulecolor{light-gray}
\toprule[0.5pt]
$L_z$ & $\tau$ & $u_0$ & $\omega_f$ & $E_{x, min}^{\tiny {\mbox{(lin)}}}/u_0$ & $E_{x, max}^{\tiny {\mbox{(0)}}}/u_0$ \\
(lbu) & (lbu) & (lbu) & (lbu) \\
\arrayrulecolor{black}
\midrule[0.5pt]
\addlinespace[0.5mm]
128 & 1.0 & $10^{-2}$ & $10^{-7}$& $0.7461 \cdot 10^{-2}$ & $0.2442 \cdot 10^{-2}$  \\
128 & 1.0 & $10^{-2}$ & $10^{-6}$& $0.7527 \cdot 10^{-2}$ & $0.6827 \cdot 10^{-2}$  \\
128 & 1.0 & $10^{-2}$ & $10^{-5}$& $1.2556 \cdot 10^{-2} $ & $1.1707 \cdot 10^{-2}$ \\
128 & 1.0 & $10^{-2}$ & $10^{-4}$& $1.8881 \cdot 10^{-2}$ & $9.9081 \cdot 10^{-2}$ \\
64 & 1.0 & $10^{-3}$ & $10^{-7}$& $1.542 \cdot 10^{-2}$ & $0.063 \cdot 10^{-2}$ \\
64 & 1.0 & $10^{-3}$ & $10^{-6}$& $1.543 \cdot 10^{-2}$ & $0.595 \cdot 10^{-2}$ \\
64 & 1.0 & $10^{-3}$ & $10^{-5}$& $1.633 \cdot 10^{-2}$ & $1.171 \cdot 10^{-2}$  \\
64 & 1.0 & $10^{-3}$ & $10^{-4}$& $2.909 \cdot 10^{-2}$ & $4.833 \cdot 10^{-2}$ \\
128 & 0.7 & $10^{-3}$ & $10^{-7}$& $0.447 \cdot 10^{-2}$ & $0.679 \cdot 10^{-2}$ \\
128 & 0.7 & $10^{-3}$ & $10^{-6}$& $0.486 \cdot 10^{-2}$ & $0.701 \cdot 10^{-2}$ \\
128 & 0.7 & $10^{-3}$ & $10^{-5}$& $5.087 \cdot 10^{-2}$ & $1.106 \cdot 10^{-2}$ \\
128 & 0.7 & $10^{-3}$ & $10^{-4}$& $12.972 \cdot 10^{-2}$ & $1.558 \cdot 10^{-2}$ \\
128 & 1.0 & $10^{-3}$ & $10^{-7}$& $0.243 \cdot 10^{-2}$ & $0.803 \cdot 10^{-2}$\\
128 & 1.0 & $10^{-3}$ & $10^{-6}$& $0.660 \cdot 10^{-2}$ & $0.810 \cdot 10^{-2}$ \\
128 & 1.0 & $10^{-3}$ & $10^{-5}$& $1.185 \cdot 10^{-2}$ & $1.192 \cdot 10^{-2}$ \\
128 & 1.0 & $10^{-3}$ & $10^{-4}$& $9.900 \cdot 10^{-2}$ & $1.893 \cdot 10^{-2}$ \\
\arrayrulecolor{light-gray}
\bottomrule[1pt]
\end{tabular}
\label{tab:parameter_space}
\end{table}

whose real part is denoted by $v_x^{\tiny \mbox{0}}(z, t)$. The velocity profile $v_x(z, t)$ has a linear limit, which is given by the penetration depth $\delta$ and the channel width  $L_z$. If $L_z / \delta \ll 1$ the condition for a linear profile is fulfilled, and we get

\begin{equation}
\label{eq:linearised}
v_x^{\tiny \mbox{(lin)}}(z, t) = u_0 \, \cos (\omega t) \left ( 1-\frac{z}{L_z}   \right ).
\end{equation}

Thus we can find an upper bound for the frequency 
\begin{equation}
\label{eq:critical_omega}
\omega_c \sim \frac{\nu}{L_z^2}
\end{equation}
so that $L_z / \delta_c \sim 1$, with $\delta_c$ being the critical penetration depth of the system. With the analytical solution at hand we can now test our LBM scheme for an external shear flow in a channel including an exact time dependence and perform  a ``scanning'' of the parameter space. In the following we define two error functions based on an $\text{L}^2$-norm. Deviations from the exact analytical solution  $v_x^{\tiny \mbox{0}}(z, t)$ are given by

\begin{equation}\label{eq:error}
E_{x}^{\tiny \mbox{(0)}} = \left [ \frac{1}{L_x L_z} \, \int_{0}^{L_x} dx \int_{0}^{L_z}  dz \, (v_x(x, z, t) - v_x^{\tiny \mbox{0}}(z, t))^2 \right ]^{\frac{1}{2}}.
\end{equation}

Moreover we define an error function with respect to the linearised solution $v_x^{\tiny \mbox{(lin)}}$

\begin{equation}\label{eq:errorlin}
E_{x}^{\tiny \mbox{(lin)}} = \left [ \frac{1}{L_x L_z} \, \int_{0}^{L_x} dx \int_{0}^{L_z}  dz \, (v_x(x, z, t) - v_x^{\tiny \mbox{(lin)}}(z, t))^2 \right ]^{\frac{1}{2}}.
\end{equation}

\begin{figure}[!htbp]
\centering
\hcenteredhbox{\includegraphics[scale=0.75]{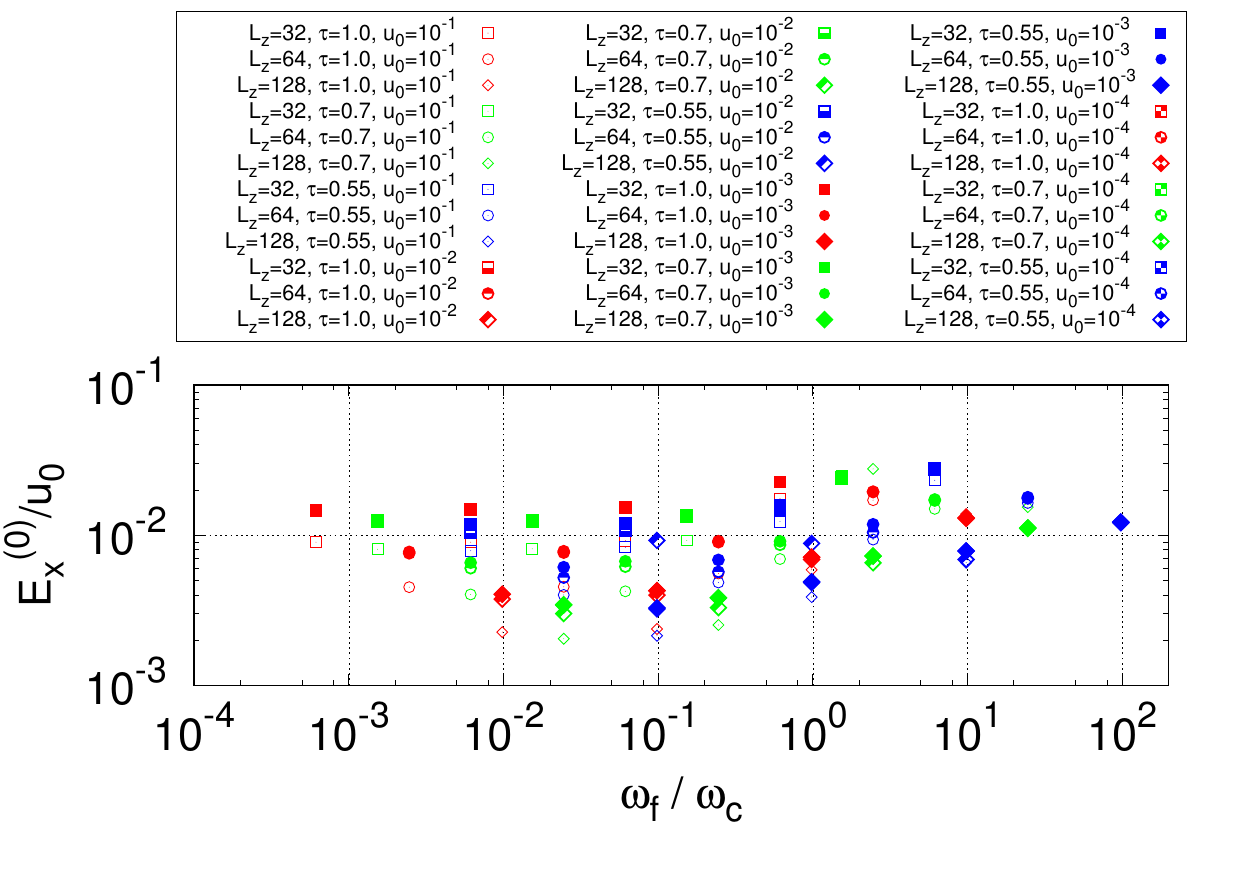}}
\hcenteredhbox{\includegraphics[scale=0.75]{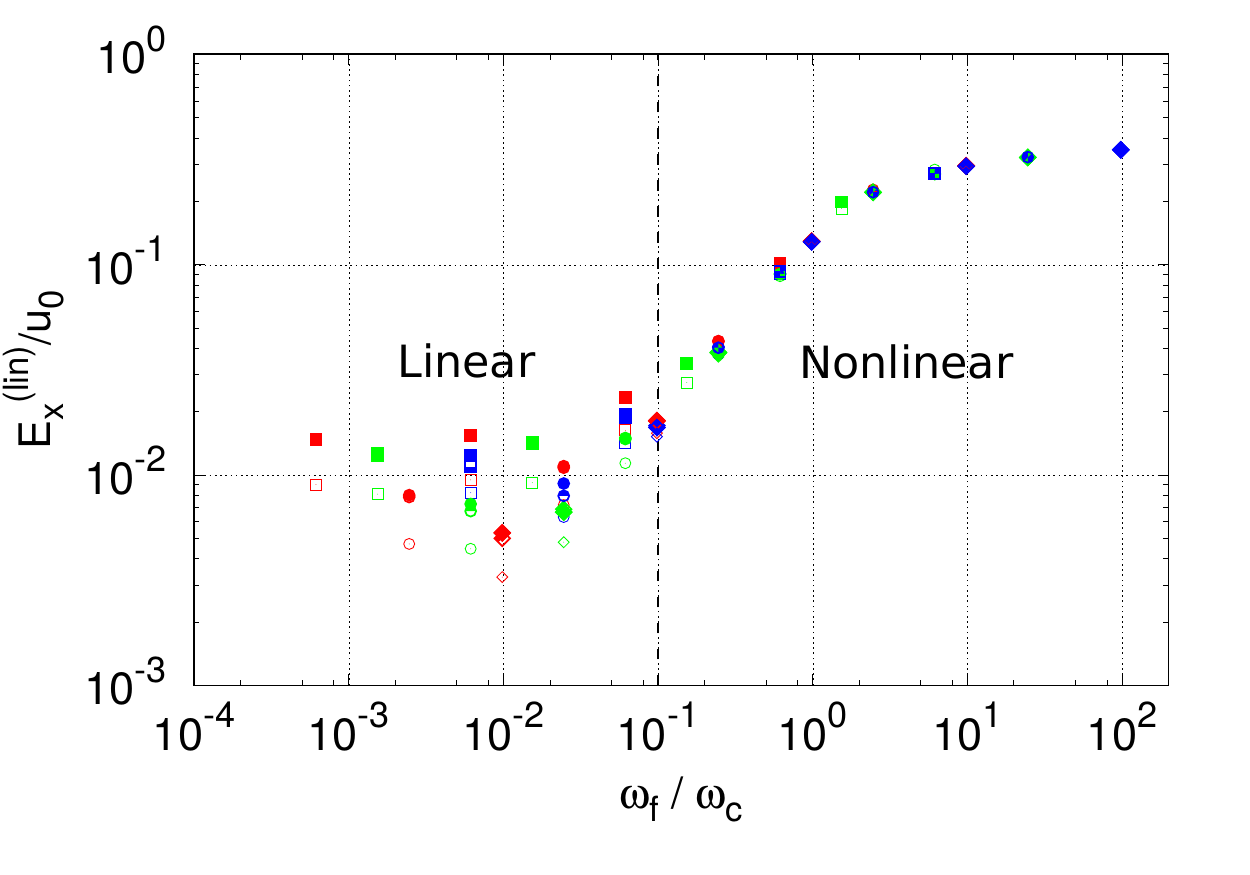}}
\caption{Scatter plot of the averaged analytical error function $E_{x}^{\tiny \mbox{(0)}} \approx \frac{1}{2 u_0} (E_{x, max}^{\tiny \mbox{(0)}} + E_{x, min}^{\tiny \mbox{(0)}})$ and averaged linear error function $E_{x}^{\tiny \mbox{(lin)}} \approx \frac{1}{2 u_0} (E_{x, max}^{\tiny \mbox{(0)}} + E_{x, min}^{\tiny \mbox{(0)}})$ as a function of the renormalised frequency $\frac{\omega_f}{\omega_c}$ with $\omega_c = \frac{\nu}{L_z^2}$ and $\omega_f \equiv \frac{\omega}{2 \pi}$. The analytical error $E_{x}^{\tiny \mbox{(0)}}$ is relatively well behaved with a mean value of around $10^{-2}$ when averaged over the whole frequency range for all parameter set-ups. On the other  hand the linear error $E_{x}^{\tiny \mbox{(lin)}}$ shows a clear dependence on the normalised frequency $\omega_f / \omega_c$. The linear error is well behaved up until a value of $\omega_f / \omega_c \sim 10^{-1}$, where the shear velocity profile starts becoming non linear, which is demonstrated by the drastic increase in $E_{x}^{\tiny \mbox{(lin)}}$. The two regions, linear and nonlinear, are separated by a dashed vertical line at $\omega_f / \omega_c = 10^{-1}$.}
\label{fig:error_collapse}
\end{figure}

We perform several simulations with different oscillation frequencies ranging from $\omega_f = 10^{-7}$ to $\omega_f = 10^{-4}$. We can estimate the critical frequency $\omega_c \approx 10^{-5}$ via eq.~(\ref{eq:critical_omega}), which is supported by fig.~\ref{fig:time_series}, as the velocity profile starts to become linear at around this frequency value. We can also see that the error $E_{x}^{\tiny \mbox{(0)}}$ with respect to the exact solution is (almost) independent of the oscillation frequency $\omega_f$. In addition, we now like to know whether our LBM simulations produce similar results for a choice of different system parameters. A sample of the parameter space is shown in table~\ref{tab:parameter_space}. Figure~\ref{fig:error_collapse} is key in understanding the validity of our LBM simulations for the single phase  oscillatory shear flow. We can see both that the exact analytical error $E_{x}^{\tiny \mbox{(0)}}$ is generally well behaved (fluctuations around a mean value) for the entire frequency range and that the error to the linearised solution $E_{x}^{\tiny \mbox{(lin)}}$ is well behaved for a frequency range $\omega_f / \omega_c \leq 10^{-1}$ and is increasing for higher frequencies. Thus, we may determine a frequency threshold of about $\omega_f / \omega_c \approx 10^{-1}$ for the linear shear regime for which our LBM solution is both stable and linear.

\section{Multicomponent oscillating flow}
\label{sec:droplet_time}

\begin{figure}[!htbp]
\vcenteredhbox{\includegraphics[scale=0.48]{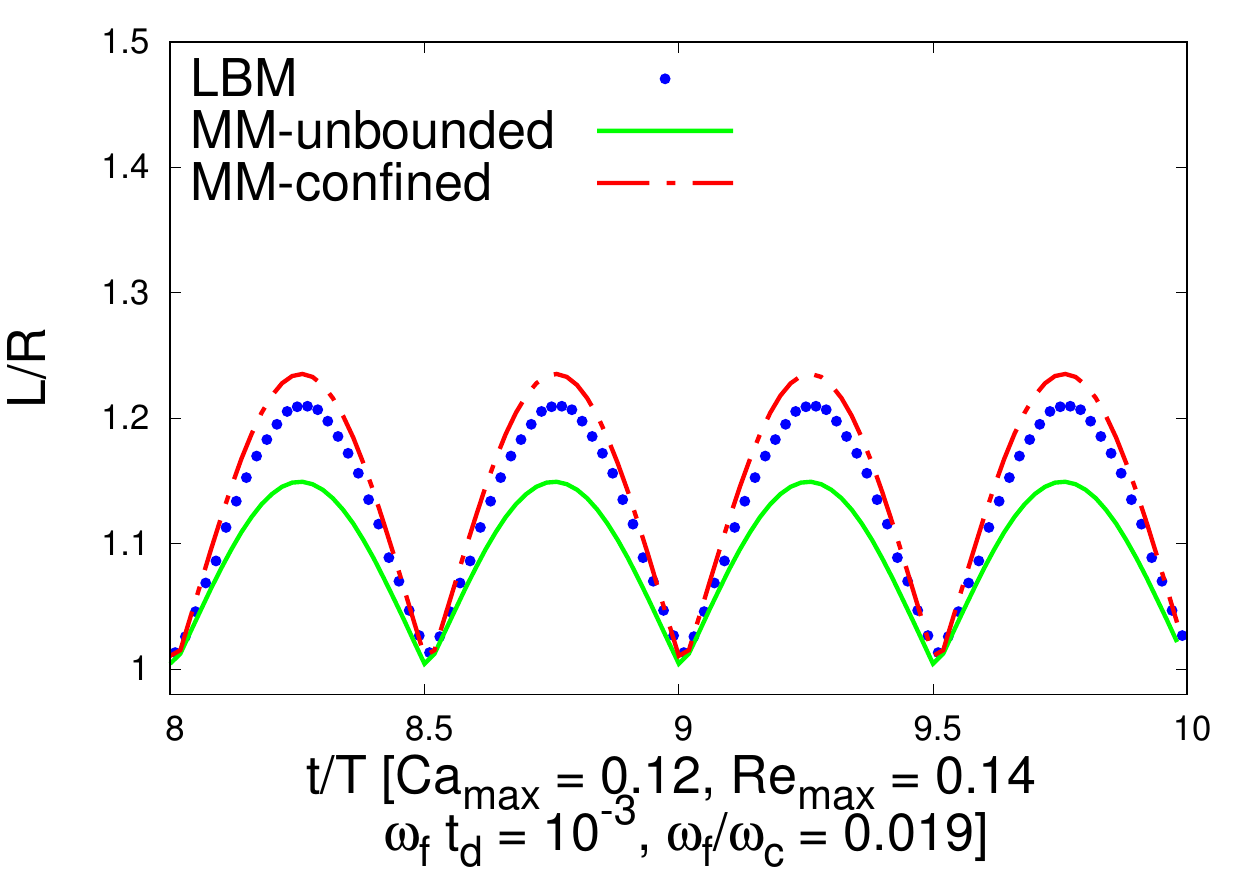}} 
\vcenteredhbox{\includegraphics[scale=0.48]{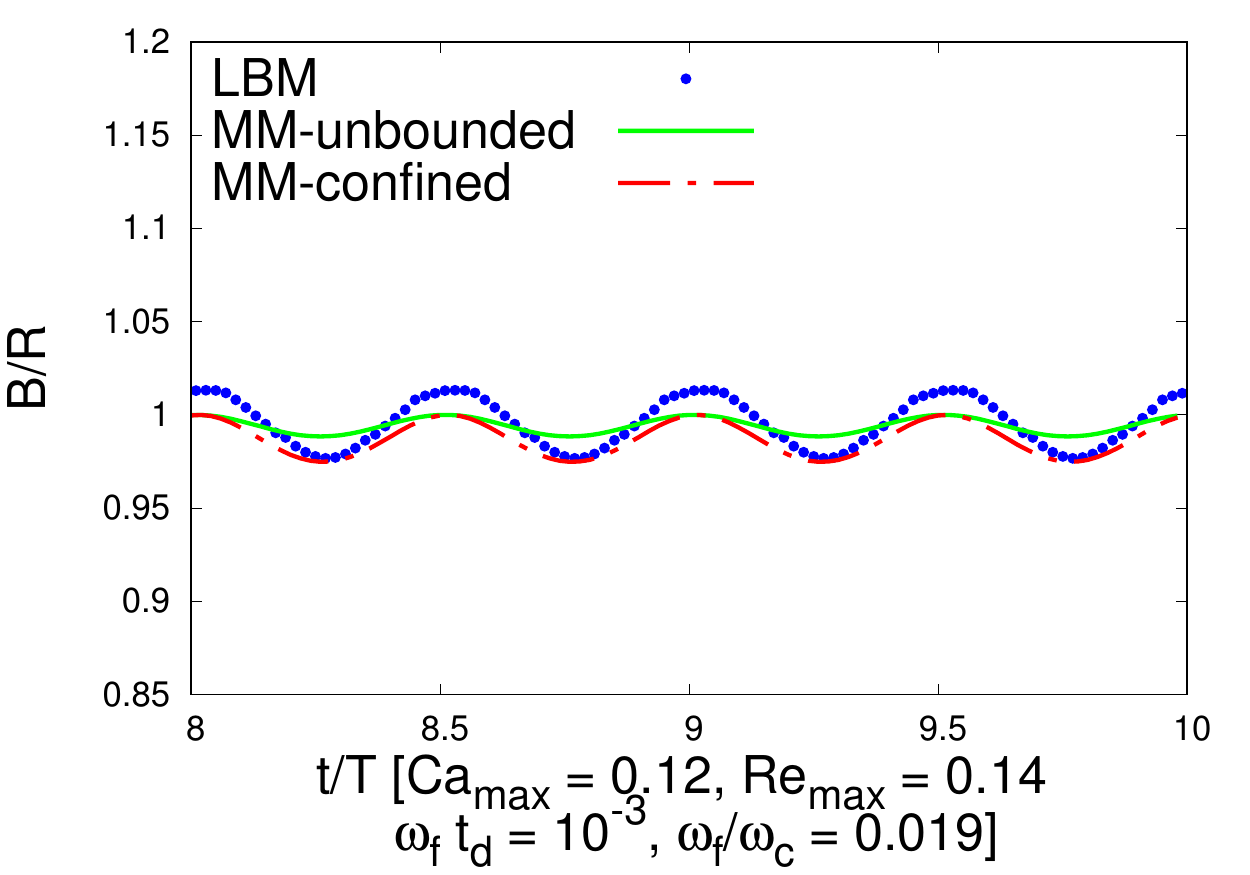}}
\vcenteredhbox{\includegraphics[scale=0.48]{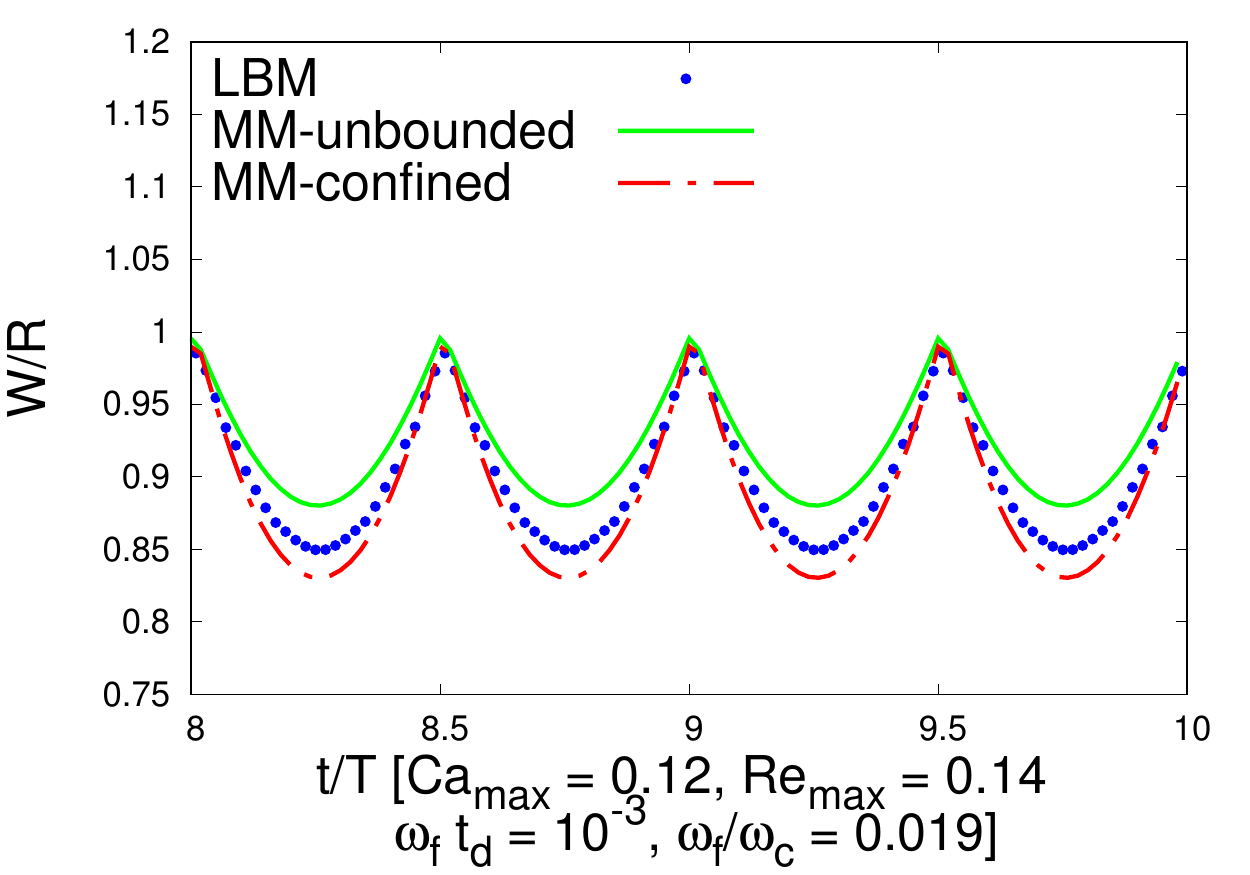}}\\
\vcenteredhbox{\includegraphics[scale=0.48]{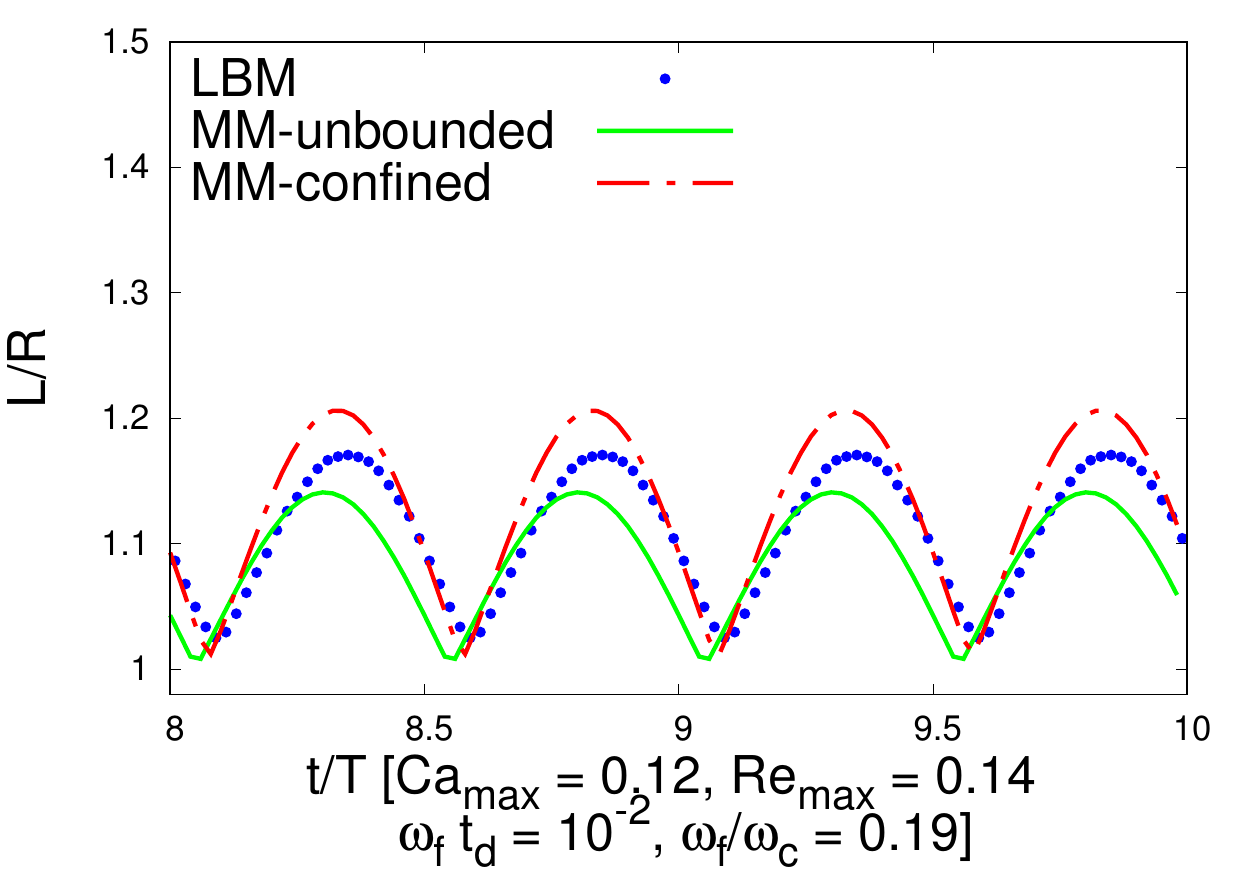}} 
\vcenteredhbox{\includegraphics[scale=0.48]{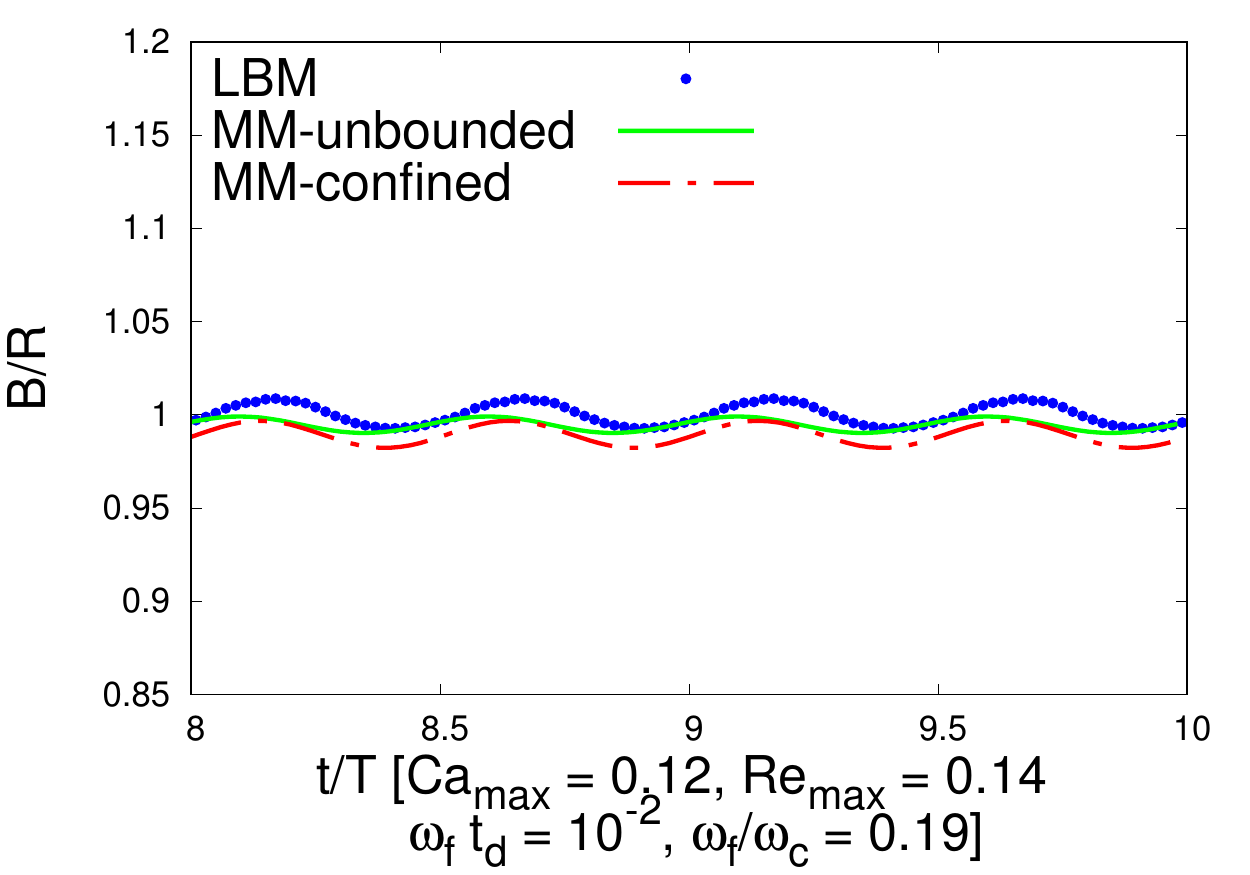}}
\vcenteredhbox{\includegraphics[scale=0.48]{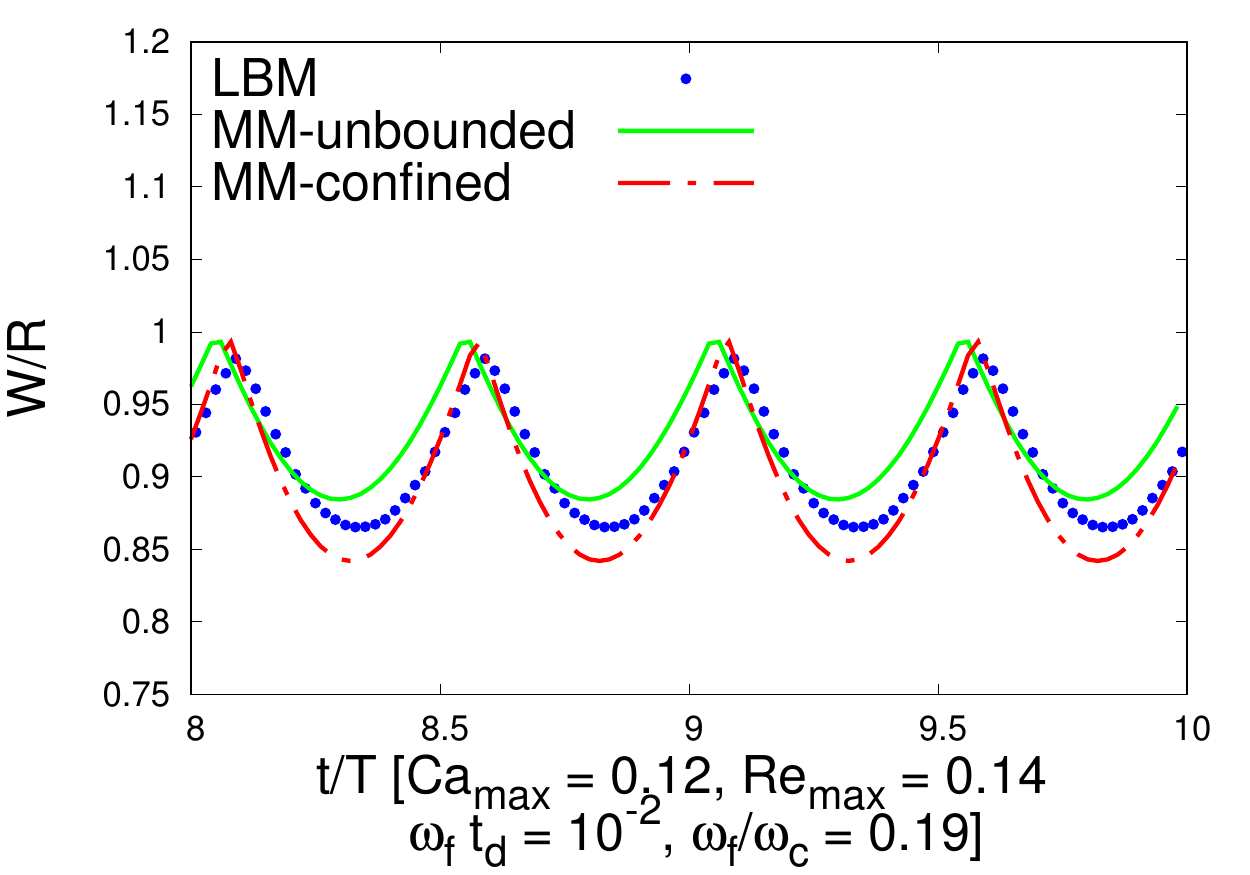}} \\
\vcenteredhbox{\includegraphics[scale=0.48]{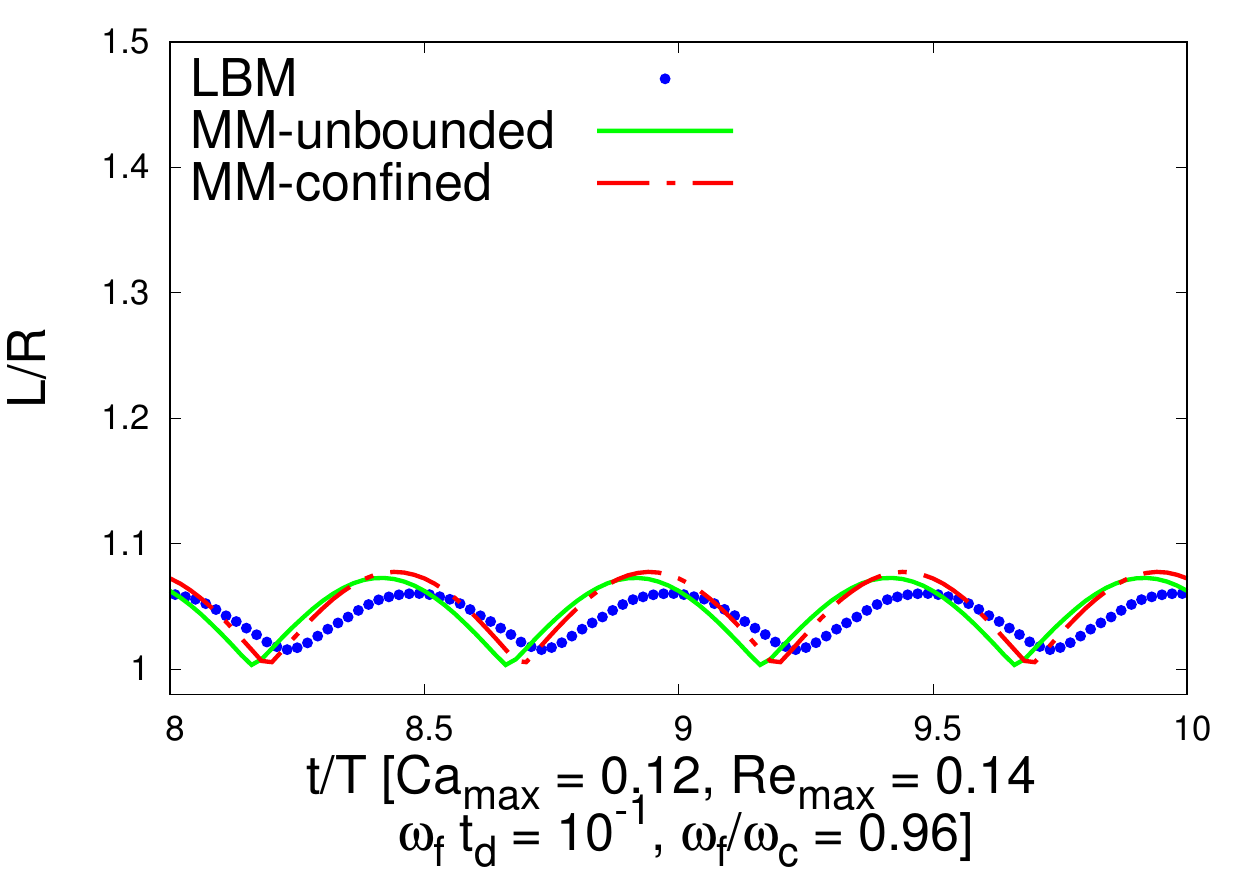}}
\vcenteredhbox{\includegraphics[scale=0.48]{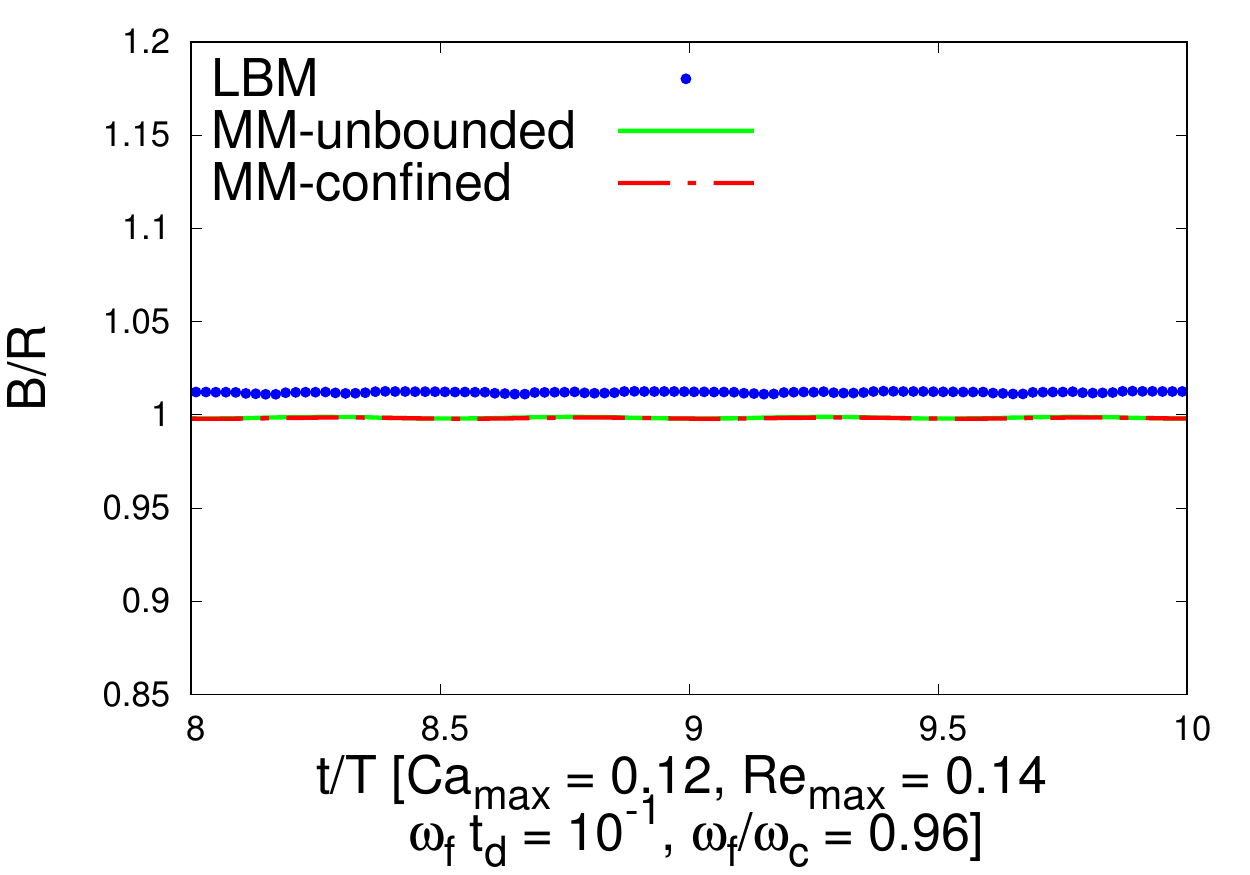}}
\vcenteredhbox{\includegraphics[scale=0.48]{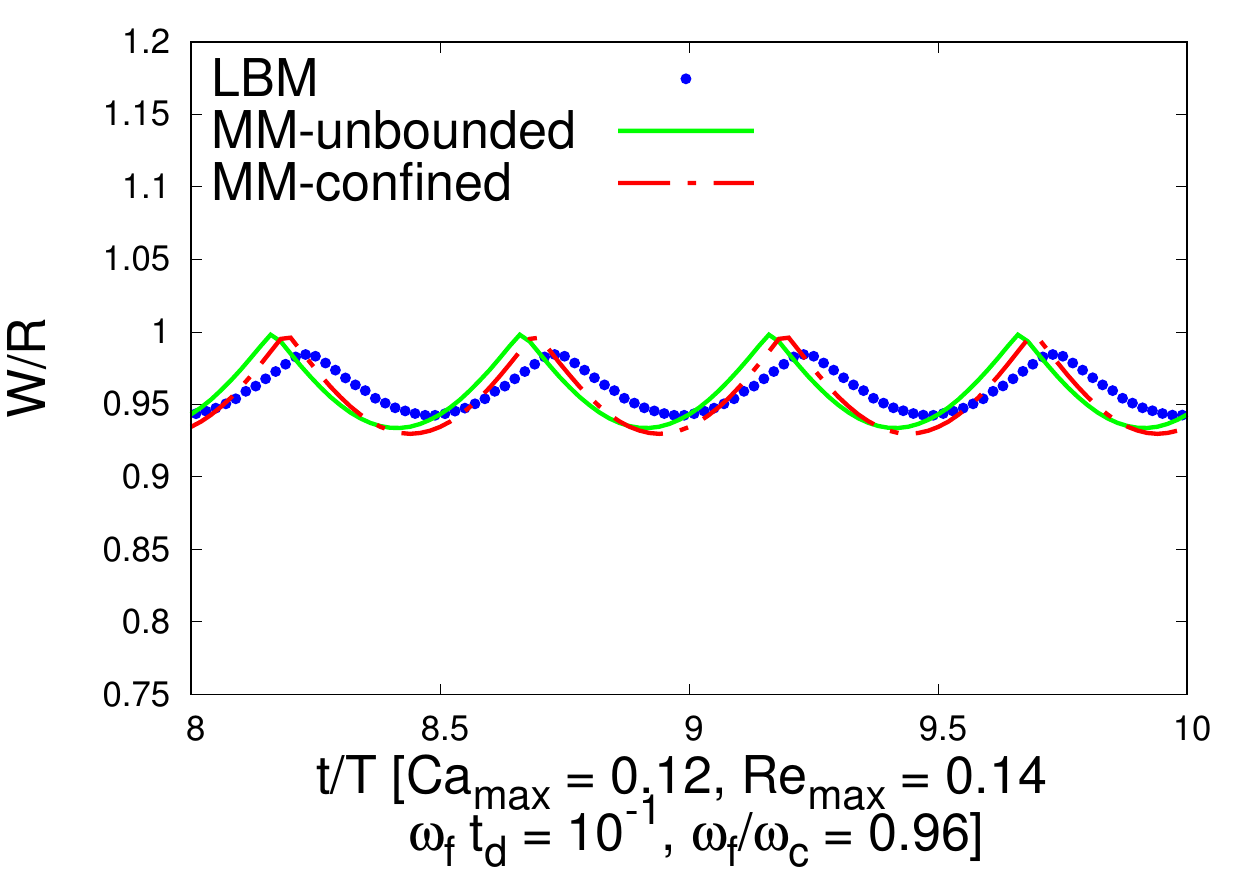}} 
\caption{Numerical benchmark for LBM against the MM-confined solution of eq.~(\ref{eq:mm_time}). The time $t$ is given in units of the shear period $T = 1 / \omega_f$. $W$ and $L$ denote the minor and major axes respectively with $B$ being the vorticity axis, where the ellipsoidal axes obey $L > B > W$ at maximal deformation. Four system parameters are of particular relevance: $\mbox{Ca}_{\text{max}}$ and $\mbox{Re}_{\text{max}}$ denote the maximal capillary and Reynolds number (given for the maximum shear at the channel walls) which remain fixed in the plots. $\omega_f / \omega_c$ is a measure of the linearity of the shear flow, where $\omega_f / \omega_c \sim 1$ may be seen as a limiting value for linearity (see sect.~\ref{sec:single_phase}). $\omega_f t_d$ denotes the oscillation frequency in units of the reciprocal droplet relaxation time and is the control parameter here. We may observe qualitatively that as the oscillation frequency tends to values close to $\omega_f t_d \sim 1$ the droplet deformation decreases and undergoes a phase shift with respect to the outer shear flow at the walls.}
\label{fig:anti_mm}
\end{figure}

After having investigated the parameter space for a single phase system, we add the droplet. We follow the discussion in \cite{Yu02} yielding a perturbative solution in $\mbox{Ca}$. Firstly we consider the MM equation in a different non-dimensionalised form with respect to eq.~(\ref{eq:mm_general}):

\begin{align}
\label{eq:mm_time}
\frac{d M_{ij}}{dt} & = \mbox{Ca} (t) \left [ f_2 (S_{ik} M_{kj} + M_{ik} S_{kj}) + \Omega_{ik} M_{kj} - M_{ik} \Omega_{kj} \right ] \notag \\
& - f_1 \left ( M_{ij} - 3 \frac{III_M}{II_M} \delta_{ij} \right ),
\end{align}

where the time $t$ is given in units of the droplet relaxation time $t_d$. It is important to note that $\mbox{Ca}(t)$ is now time-dependent due to the time-dependent external shear flow \cite{Yu02,Cox69,Cavallo02,Farutin2012}. Following the discussion in \cite{Yu02} we will expand the morphology tensor $M_{ij}$ as a perturbation series in the capillary number $\mbox{Ca}(t)$. Ignoring an initial transient, we end up with the following first order solutions for the squared ellipsoidal axes

\begin{align}
\label{eq:mm_eigen_real}
\hat{L}^2 & = 1 + \mbox{Ca}_{\text{max}} f_2 \left ( \frac{\omega t_d \cos ( \omega  t_d t) - f_1 \sin (\omega t_d t)}{f_1^2 + \omega^2 t_d^2} \right ) \notag \\
& + \mathcal{O}(\mbox{Ca}_{\text{max}}^2), \notag \\
\hat{W}^2 & = 1 - \mbox{Ca}_{\text{max}} f_2 \left ( \frac{\omega t_d \cos ( \omega t_d t) - f_1 \sin (\omega t_d t)}{f_1^2 + \omega^2 t_d^2} \right ) \notag \\
& + \mathcal{O}(\mbox{Ca}_{\text{max}}^2), \notag \\
\hat{B}^2 & = 1 + \mathcal{O}(\mbox{Ca}_{\text{max}}^2),
\end{align}

where $\mbox{Ca}_{\text{max}}$ denotes the maximal capillary number and $t$ is given in units of $t_d$. The quantities $L^2$, $B^2$ and $W^2$ denote the maximal, medium and minimal eigen-direction of the morphology tensor $M_{ij}$ at all times $t$ and are defined via

\begin{align}
\label{eq:norm_max}
L^2 & = \lvert \lvert \hat{L}^2, \hat{W}^2 \rvert \rvert_{\infty}, \notag \\
W^2 & = 2 - \lvert \lvert \hat{L}^2, \hat{W}^2 \rvert \rvert_{\infty}, \notag \\
B^2 & = \hat{B}^2,
\end{align}

where $\lvert \lvert a, b \rvert \rvert_{\infty} \equiv \text{max} (\lvert a \rvert ,\lvert b \rvert)$ is the maximum norm between two scalar quantities $a$ and $b$. Besides the three ellipsoidal axes $L$, $B$ and $W$ another quantity is of particular interest to us. Analogously to \cite{Yu02} we applied a sinusoidal shear rate, so that for the time-dependent capillary number $\mbox{Ca} (t) \sim \sin (\omega t_d t)$ ($t$ in units of $t_d$). Thus we can identify a phase shift $\phi$ between the external oscillatory shear and the droplet's response given by the time evolution of the squared ellipsoidal axes in eq.~(\ref{eq:mm_eigen_real}):

\begin{equation}
\label{eq:phase_shift}
\phi = \arctan \left ( \frac{\omega t_d}{f_1} \right ) + \mathcal{O} (\mbox{Ca}_{\text{max}}^2),
\end{equation}

which is (for the linearised solution) independent of $\mbox{Ca}_{\text{max}}$. With our theoretical model at hand we can now run LBM simulations of the droplet in the oscillatory shear flow and check the agreement with the perturbative theoretical predictions in eq.~(\ref{eq:mm_eigen_real}). However, the perturbative analytical solution is only valid in a small capillary $\mbox{Ca}_{\text{max}}$ range. We can also solve the time-dependent MM-confined eq.~(\ref{eq:mm_time}) and compare the numerical solution (obtained via a RK-4 scheme) to our LBM simulation results instead of the perturbative solution. It should be remarked that our LBM simulation results may only be compared to the MM-confined model, when the droplet remains an ellipsoid at all times. In order to have a complete overview of the droplet deformation it is paramount to visualise all three major ellipsoidal axes $L$, $B$, $W$, where $L > B > W$ at maximum deformation. Let us look at the top row of fig.~\ref{fig:anti_mm}. Considering the time evolution of the major axis $L$ and minor axis $W$ we can see that MM-unbounded model is not properly accounting for the confinement of the system ($\alpha = 0.75$). On the other hand our LBM simulation results are in relatively good agreement with the numerical solution of MM-confined. Interestingly, the vorticity axis $B$ is also deformed in time, which is not the case in the perturbative model, since the deformation is due to higher orders $\mathcal{O}(\mbox{Ca}_{\text{max}}^2)$ in this case. Moving one row further down in fig.~\ref{fig:anti_mm}, i.e. increasing the previous frequency by a factor $10$ we can observe two changes. Firstly, the value of the droplet deformation $D$ is decreasing (for both LBM and the MM-confined solution), and secondly, the time evolution is shifted with respect to the previous row. These effects may be explained by the droplet inertia which tries to resist the outer shear flow. Since $\omega_f t_d \sim 10^{-2}$ we are in the regime where the droplet

\onecolumn
\twocolumn

\begin{figure}[htbp]
\centering
\includegraphics[scale=0.75]{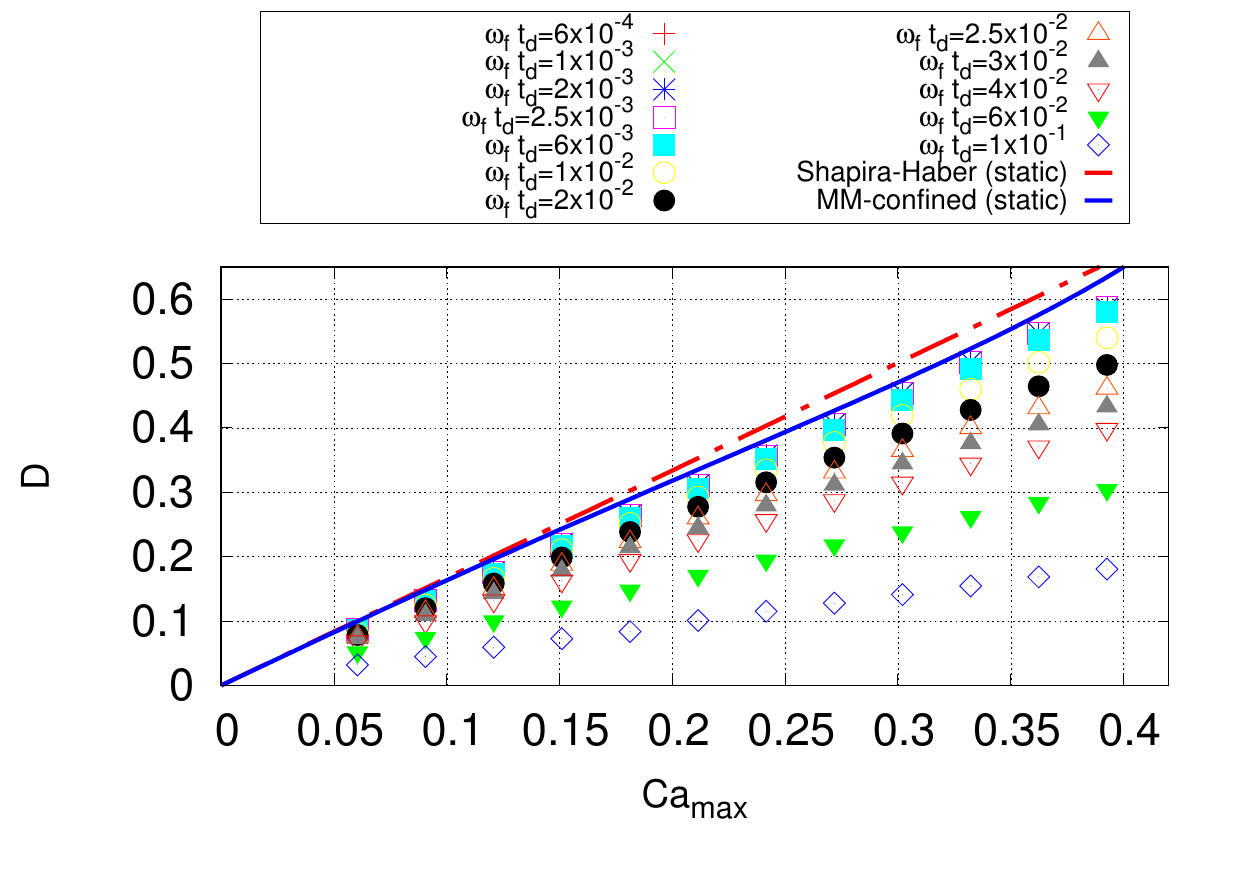}
\caption{LBM obtained droplet deformation $D$ against $\mbox{Ca}_{\text{max}}$ for various values of the normalised shear frequency $\omega_f t_d$. The transparency effect, i.e. the reduction of deformation $D$ for $\omega_f t_d \to 1$ (see eq.~(\ref{eq:mm_eigen_real}) and fig.~\ref{fig:anti_mm}) is confirmed. For further clarification the Shapira-Haber and the MM-confined static deformation curves are shown, from which we may see that a relatively low oscillation frequency of about $\omega_f t_d \sim 10^{-3}$ results already in a noticeable decrease in $D$.}
\label{fig:droplet_phase_defo}
\end{figure}

\begin{figure}[htbp]
\centering
\includegraphics[scale=0.75]{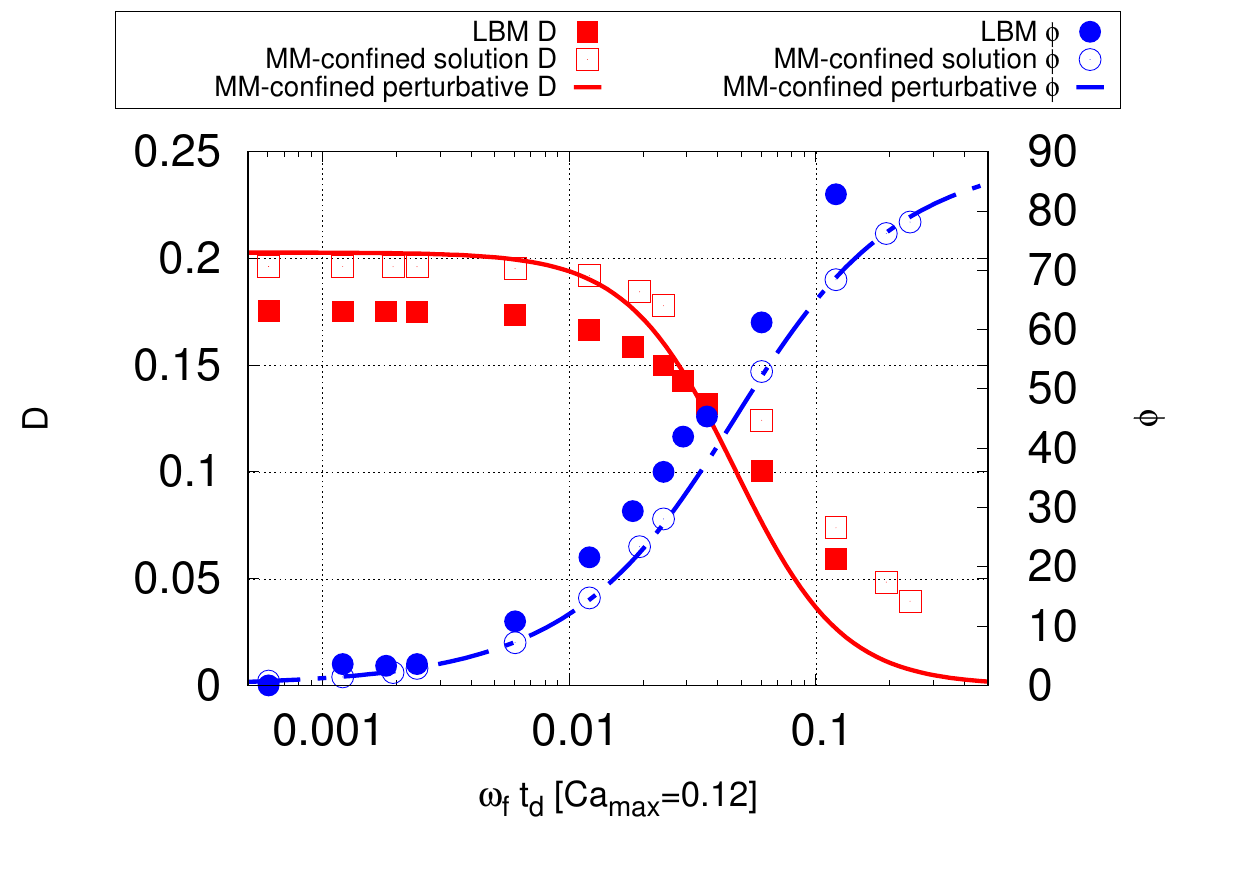}
\caption{Droplet deformation $D$ and phase shift $\phi$ between the shear flow and the droplet's response for a fixed maximal capillary number $\mbox{Ca}_{\text{max}}$ against the normalised frequency $\omega_f t_d$. The LBM measured droplet deformation $D$ is in good agreement with the MM-confined prediction (see fig.~\ref{fig:droplet_static} for the static benchmark). The droplet transparency effect seems to come into effect at around $\omega_f t_d \sim 10^{-2}$, where a gradual decrease in $D$ is noticeable in both MM-confined and LBM results. MM-confined and LBM agree furthermore on the phase shift $\phi$ which increases from $\phi \sim 0$ to $\phi \sim \pi / 2$ for the highest measured frequencies. This indicates an out of phase  droplet response to the underlying shear flow. For further clarification the linearised perturbative MM-confined solution, eq.~(\ref{eq:mm_eigen_real}), is also shown.}
\label{fig:droplet_phase_omega}
\end{figure}

relaxation time scale is relatively close to the oscillatory shear period $1 / \omega_f$. Therefore, we may expect both a decrease in deformation and a phase shift $\phi$ between the outer shear flow and the time-dependent droplet deformation $D(t)$, as is also predicted by the analytical perturbative solution (see eq.~(\ref{eq:phase_shift})). The phase shift $\phi$ is measured in the LBM simulations by the difference in simulation time between the maximal shear intensity $G$ and the maximal droplet deformation $D$. Increasing the frequency even more to $\omega_f t_d \approx 10^{-1}$ the deformation decreases substantially and the phase shift $\phi$ is close to $\pi / 2$. This indicates that as $\omega_f t_d \to 1$ the droplet is behaving as if the flow was not present at all. We call this the ``transparency'' effect, since the droplet seems to be (almost) transparent to the surronding flow field, which makes itself noticeable by the droplet's out of phase response and drastic decrease in the deformation parameter $D$. This decrease in deformation due to a phase shift between applied shear and droplet response has also been experimentally confirmed by Cavallo et al. \cite{Cavallo02}, where the authors use a different small amplitude model as a benchmark for their experimental results. For further analysis fig.~\ref{fig:droplet_phase_defo} shows the LBM droplet deformation results as a function of $\mbox{Ca}_{\text{max}}$ for the simulated frequency range. In fig.~\ref{fig:droplet_phase_defo} the transparency effect is shown in a more quantitative way. We observe for various simulations, that the deformation drops significantly for increasing frequency $\omega_f$, independently of the capillary number $\mbox{Ca}$. For further comparison of the droplet deformation scale $D$ the Shapira-Haber \cite{ShapiraHaber90} and MM-confined curves \cite{Minale08} are given as well. Figure~\ref{fig:droplet_phase_omega} shows both the deformation $D$ and the phase shift $\phi$ between the droplet response and the oscillatory shear flow as a function of the normalised frequency $\omega_f t_d$.  The general trend is that the deformation $D$ is stationary up until $\omega_f t_d \sim 0.01$ at which point $D$ starts decreasing until the droplet becomes ``transparent" to the outer shear flow. We may also see 

\begin{figure}[!htbp]
\centering
\textbf{Low non-dimensionalised shear frequency} $\mathbf{\omega_f t_d = 0.001}$ \textbf{for} $\mathbf{Ca_{\textbf{max}} = 0.12}$
\vcenteredhbox{\includegraphics[scale=0.25, trim={100mm, 80mm, 100mm, 20mm},clip]{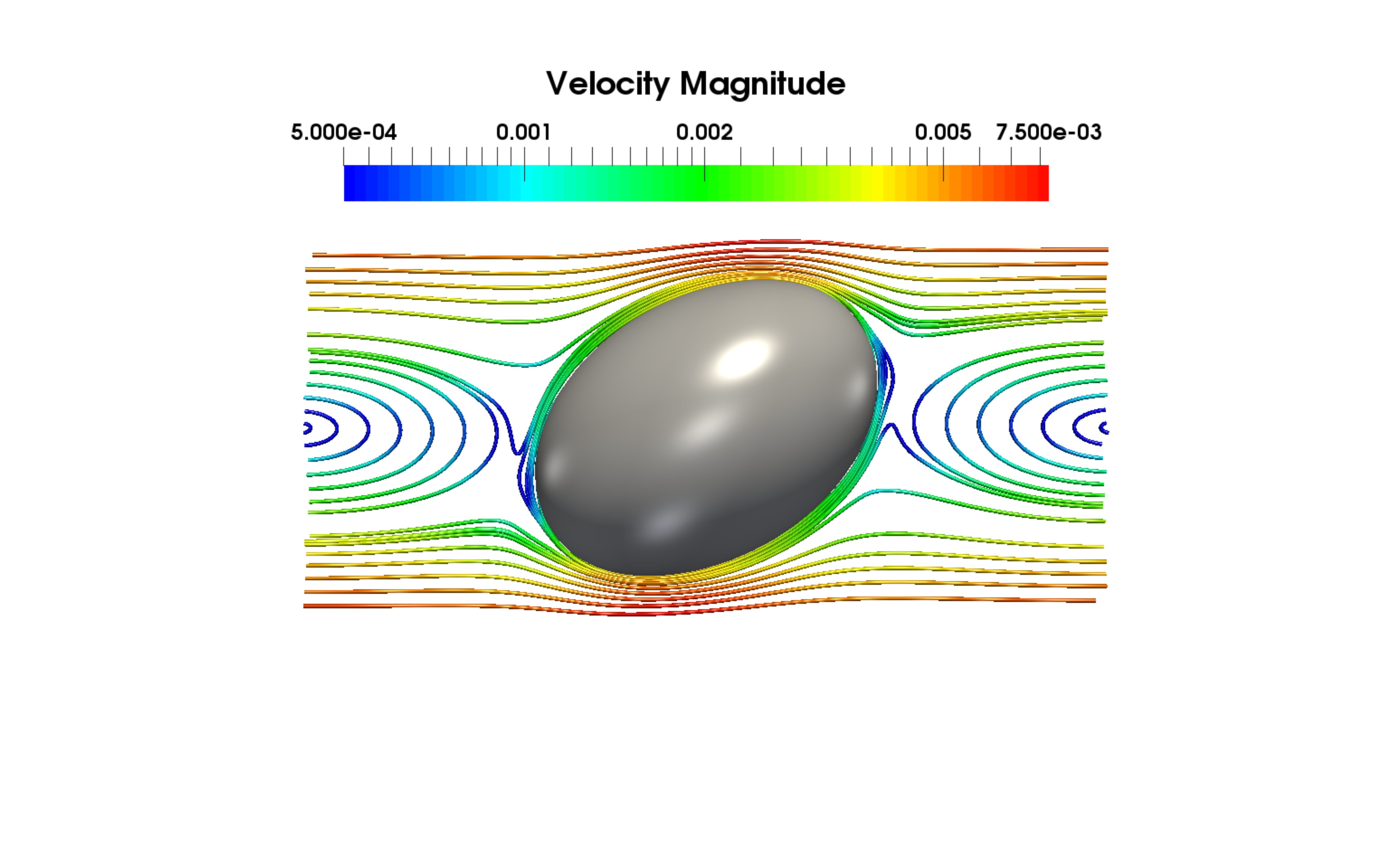}}
\textbf{High non-dimensionalised shear frequency} $\mathbf{\omega_f t_d = 0.1}$ \textbf{for} $\mathbf{Ca_{\textbf{max}} = 0.12}$
\vcenteredhbox{\includegraphics[scale=0.25, trim={100mm, 80mm, 100mm, 20mm},clip]{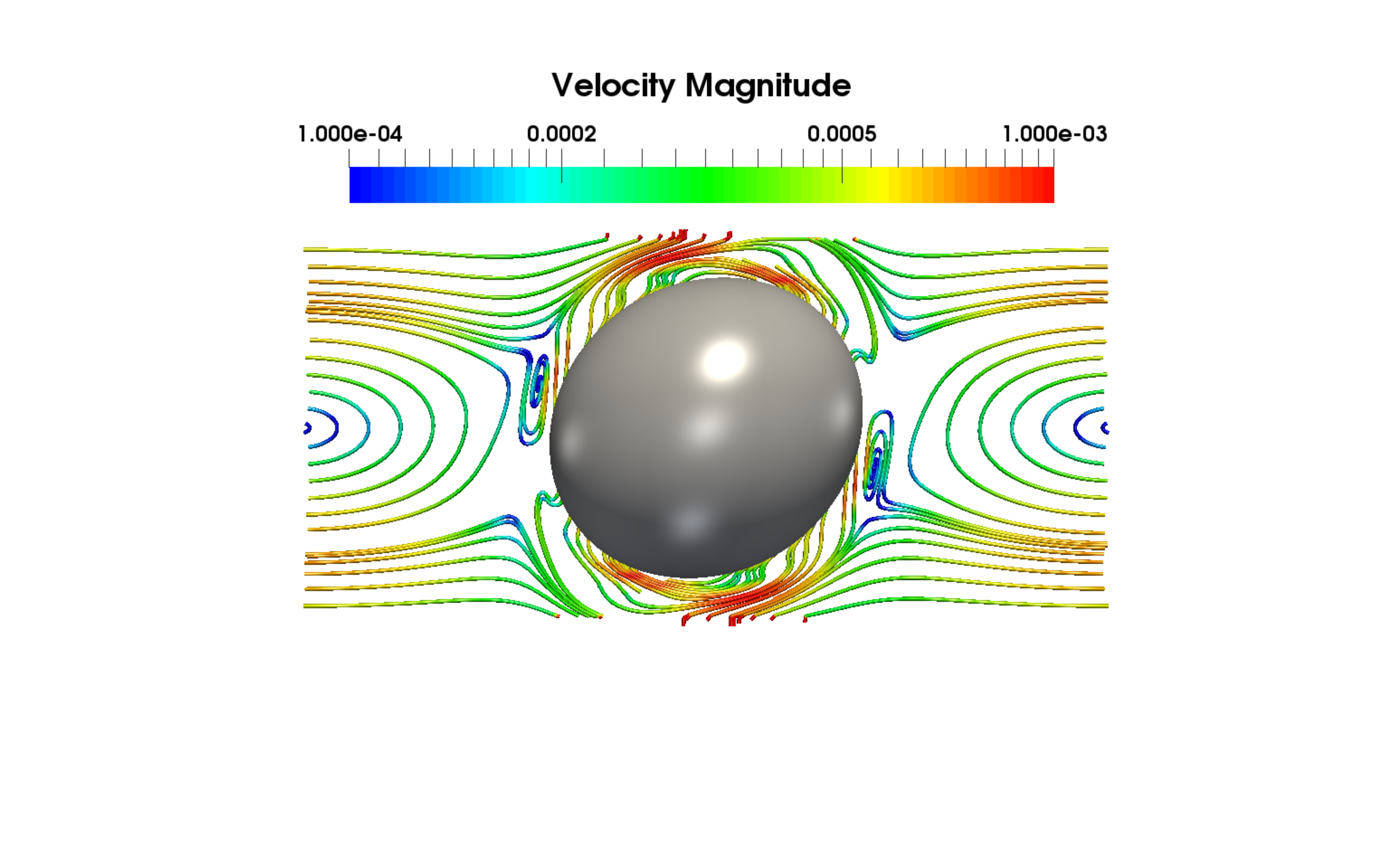}}
\caption{Streamline plots of the LBM droplet simulations at maximum deformation for $\mbox{Ca}_{\text{max}} = 0.12$ and $\mbox{Re}_{\text{max}} \approx 0.1$. Top panel: Low frequency regime $\omega_f t_d = 0.001$. The droplet is ellipsoidally deformed and tilted in the channel, similarly to static droplet deformation dynamics \cite{GuptaSbragaglia2014}. The droplet produces two wakes in the channel and the velocity magnitude is largest at the channel walls. Bottom panel: High frequency regime $\omega_f t_d = 0.1$. The droplet is only marginally deformed due to the ``transparency effect" at non-dimensionalised high frequencies $\omega_f t_d$. We observe once again two wakes in the flow field in the vicinity of the droplet. Due to $\phi \approx \pi / 2$ (phase shift between underlying oscillatory shear and droplet deformation) the velocity magnitude is largest at the droplet interface instead of the channel walls. The droplet is deforming the underlying oscillatory shear flow through its own internal dynamics (two-way coupling).}
\label{fig:droplet_streamlines}
\end{figure}

that the phase shift $\phi$ is starting to increase rapidly at $\omega_f t_d \sim 0.01$  from $\phi \sim 0$ up to $\phi \sim \pi / 2$. This reinforces the idea of the droplet transparency effect. Similarly to a forced harmonic oscillator the shear flow is out of phase with the droplet's response, because the oscillatory shear period $1 / \omega_f$ is of comparable size to the droplet relaxation time $t_d$. The numerical solution of the MM-confined eq.~(\ref{eq:mm_time}) shown in fig.~\ref{fig:droplet_phase_omega} is in good agreement with the perturbative analytical solution eq.~(\ref{eq:mm_eigen_real}) for both $D$ and $\phi$.   The LBM results predict a smaller deformation $D$ and a larger phase shift $\phi$ compared to the MM-confined model. This may be explained by the very thin droplet interface in the LBM simulation which is roughly the size of $1$ grid point. From this estimation we can deduce a relative error of about $0.02$ to both the time-dependent values of $L$ and $W$, resulting in a relative error of about $0.01$ for the deformation parameter $D$. Moreover, it is useful to qualitatively consider streamline plots of the droplet dynamics in both the low and high frequency regions (see fig.~\ref{fig:droplet_streamlines}). In the low frequency regime for $\omega_f t_d = 0.001$ in fig.~\ref{fig:droplet_streamlines} we see the familiar case of static droplet deformation \cite{GuptaSbragaglia2014}, where we have a tilted ellipsoidally deformed droplet (in agreement with the MM-confined model) in the case of maximum deformation coinciding with the instance of the maximum shear due to $\phi \ll 1$. In the high frequency regime $\omega_f t_d = 0.1$ in fig.~\ref{fig:droplet_streamlines} we see now that the droplet is only slightly deformed in the case of maximal deformation. Since the phase shift $\phi \approx \pi /2$ now, the velocity magnitude of the oscillatory shear flow is almost $0$ at the walls. We observe that the regions of highest shear flow intensity are in fact close to the droplet interface (disregarding the two channel wakes produced by the droplet). Thus the internal droplet dynamics substantially influences the oscillatory shear flow close to the interface in the high frequency regime. This is a consequence of the two-way coupling of the Multicomponent LBM scheme.

\section{Conclusions and Outlook}

We have demonstrated that a Shan-Chen multicomponent LBM set-up with particularly chosen boundary conditions yields reliable results for confined time-dependent droplet deformation. After validations in the static case \cite{Taylor32,ShapiraHaber90}, we have checked the LBM results against a variety of time-dependent theoretical models \cite{Minale08,MaffettoneMinale98,Yu02}. Specifically, after introducing a time dependence into the system via a monochromatic shear, the LBM simulation agree fairly well with theoretical models and discrepancies are likely due to the interface thickness in the LBM model. The simulations in this work have been carried out with a boundary scheme using ghost nodes which is, on the one hand, equivalent to a wall bounce back scheme, but, on the other hand, may be extended to model more complex shear flows than those treated here. Therefore, our simulations both for single and multi-component flows are useful benchmarks of a boundary method involving ghost nodes, which can be extended to a pressure driven boundary scheme~\cite{Mattila09,Hecht10}. This work may also be extended to consider the rather interesting aspect of frequency dependent droplet break up which may be seen as an extension to a previous work on Reynolds number dependent droplet break up \cite{RenardyCristini01}. The aspect of time-dependent droplet break up in a simple shear flow is currently being investigated. Furthermore, it is interesting to see whether the underlying LBM boundary scheme described here may be extended to accurately simulate an ``ab-initio" droplet in a turbulent flow \cite{Njobuenwu2015,Biferale2014,Spandan2016}.\\

The authors kindly acknowledge funding from the European Union's Framework Programme for Research and Innovation Horizon 2020 (2014 - 2020) under the Marie 
Sklodowska-Curie Grant Agreement No.642069 and funding from the European Research Council under the European Community's Seventh Framework Program, ERC Grant Agreement No 339032. The authors would also like to thank Dr Fabio Bonaccorso, Dr Anupam Gupta, Xiao Xue and Gianluca di Staso for their support.

\section*{Author contribution statement}

All of the authors were involved in the preparation of the manuscript and have read and approved the final manuscript version.

\end{document}